\documentclass{article}
\usepackage{ijcai25}

\usepackage{times}
\usepackage{soul}
\usepackage{url}
\usepackage[hidelinks]{hyperref}
\usepackage[utf8]{inputenc}
\usepackage[small]{caption}
\usepackage{graphicx}
\usepackage{amsmath}
\usepackage{amsthm}
\usepackage{booktabs}
\usepackage{algorithm}
\usepackage{algorithmic}
\usepackage[switch]{lineno}

\newtheorem{theorem}{Theorem}

\usepackage{algorithm}
\usepackage{algorithmic}
\usepackage{amsfonts}
\usepackage{xcolor}
\usepackage{multirow}
\usepackage{graphicx} % for pdf, bitmapped graphics files
\usepackage{subfigure}
\usepackage{xspace}

\newtheorem{lemma}{Lemma}
\newtheorem{definition}{Definition}

\newcommand{\graphsize}{\fpeval{0.15}} % Assign a value to \myvar

\newcount\Comments  % 0 suppresses notes to selves in text
\long\def\roie#1{{\ifnum\Comments=1\color{red} [Roie: #1]\fi}}
\long\def\roni#1{{\ifnum\Comments=1\color{purple} [Roni: #1]\fi}}
\long\def\ariel#1{{\ifnum\Comments=1\color{red} [Ariel: #1]\fi}}
\long\def\arseni#1{{\ifnum\Comments=1\color{red} [Arseni: #1]\fi}}
\newcommand{\cgapibt}{MACGA+PIBT\xspace}
\newcommand{\cgamapf}{MACGA\xspace}

\newcommand{\createcorridor}{{\tt CreateCorridor}\xspace}
\newcommand{\evacuate}{{\tt Evacuate}\xspace}
\newcommand{\push}{{\tt Push}\xspace}
\newcommand{\commentout}[1]{ }
\title{Multi-Agent Corridor Generating Algorithm}

\author{
Arseniy Pertzovsky
\and
Roni Stern\and
Roie Zivan\And
Ariel Felner\\
\affiliations
Ben-Gurion University of the Negev\\
\emails
arsenip@post.bgu.ac.il,
roni.stern@gmail.com,
\{zivanr, felner\}@bgu.ac.il
}
\begin{document}

\maketitle

\begin{abstract}
In this paper, we propose the Multi-Agent Corridor Generating Algorithm (MACGA) for solving the Multi-agent Pathfinding (MAPF) problem, where a group of agents need to find non-colliding paths to their target locations. Existing approaches struggle to solve dense MAPF instances. In MACGA, the agents build \emph{corridors}, which are sequences of connected vertices, from current locations towards agents' goals, and evacuate other agents out of the corridors to avoid collisions and deadlocks. We also present the MACGA+PIBT algorithm, which integrates the well-known rule-based PIBT algorithm into MACGA to improve runtime and solution quality. The proposed algorithms run in polynomial time and have a reachability property, i.e., every agent is guaranteed to reach its goal location at some point. We demonstrate experimentally that MACGA and MACGA+PIBT outperform baseline algorithms in terms of success rate, runtime, and makespan across diverse MAPF benchmark grids.
\end{abstract}

%%%%%%%%%%%%%%%%%%%%%%%%%%%%%%%%%%%%%%%%%%%%
%%%%%%%%%%%%%%%%%%%%%%%%%%%%%%%%%%%%%%%%%%%%
%%%%%%%%%%%%%%%%%%%%%%%%%%%%%%%%%%%%%%%%%%%%
%%%%%%%%%%%%%%%%%%%%%%%%%%%%%%%%%%%%%%%%%%%%
%%%%%%%%%%%%%%%%%%%%%%%%%%%%%%%%%%%%%%%%%%%%
%%%%%%%%%%%%%%%%%%%%%%%%%%%%%%%%%%%%%%%%%%%%
%%%%%%%%%%%%%%%%%%%%%%%%%%%%%%%%%%%%%%%%%%%%
%%%%%%%%%%%%%%%%%%%%%%%%%%%%%%%%%%%%%%%%%%%%
%%%%%%%%%%%%%%%%%%%%%%%%%%%%%%%%%%%%%%%%%%%%
\section{Introduction}

% MAPF is an important problem and there are many solvers. We propose a new one that is suboptimal but super fast. 
Multi-agent Pathfinding (MAPF) is the problem of finding a set of non-colliding paths for a group of agents to their goal locations~\cite{stern2019mapf}. 
Instances of MAPF exist in robotics~\cite{bartak2019multi}, automated warehouses~\cite{wurman2008coordinating,salzman2020research}, digital entertainment~\cite{ma2017feasibility} and many more~\cite{morris2016planning}. 
Many MAPF algorithms have been proposed in the past decade~\cite{felner2017search,li2021anytime,okumura2023lacam}. 
% Some MAPF algorithms are \emph{optimal} in the sense that the solution they return is guaranteed to have the lowest cost according to some cost function, and some MAPF algorithms are \emph{complete}, which means they are guaranteed to return a solution if one exists. 
Some MAPF algorithms are \emph{complete} in the sense that they are guaranteed to return a solution if one exists, and some are also \emph{optimal}, i.e., they are guaranteed to have the lowest cost according to some cost function. 

Optimal and complete MAPF solvers struggle to solve to 
large-scale MAPF instances.
%, however, optimal and complete MAPF solvers are rarely trading off solution quality and completeness for faster runtime. 
By contrast, MAPF algorithms such as Prioritized Planning (PrP)~\cite{prp_2015}, PIBT~\cite{okumura2022priority}, MAPF-LNS2~\cite{li2022mapf}, and LaCAM*~\cite{okumura2023improving} often scale to very large problems with many agents by trading optimality and sometimes even completeness for faster running time. 
% \arseni{to mention LaCAM$^*$~\cite{okumura2023improving} algorithm here? or unnecessary?..}
% \roni{I revised the text here and above to allow us to mention LACAM* here also. The rationale is to talk about optimal and complete algorithms, and algorithms that are not optimal and complete. This way we group LACAM* and LNS2 in the same "bundle. Anyhow, please -reread this and the previous paragraph.}
Yet these algorithms often fail to solve small and dense MAPF problems. 
In this work, we propose a novel suboptimal and incomplete MAPF algorithm
called the Multi-Agent Corridor Generating Algorithm (\cgamapf), which outperforms existing algorithms on several benchmark MAPF domains, including dense small maps. 
% How it works
\cgamapf builds on the Corridor Generating Algorithm (CGA) algorithm, which is a recently introduced algorithm for the Single-Agent Corridor Generating (SACG)~\cite{arseni2024corr} problem. 
In SACG, an agent denoted as the \emph{main agent}, needs to arrive at its goal and it is allowed to push away other agents out of its way. MAPF can be viewed as a generalization of SACG, in which each agent has a goal, and they all need to arrive at their destination. 
\cgamapf aims to capture this intuition by iterating over the agents one at a time.

\cgamapf has several attractive properties. 
Unlike many MAPF algorithms, which require heavy computation (e.g., LaCAM~\cite{okumura2023lacam}, LaCAM$^*$~\cite{okumura2023improving} or CBS~\cite{CBS_2015}), \cgamapf is tractable, running in polynomial time. 
Similar to PIBT~\cite{okumura2022priority}, it guarantees \emph{reachability}, which means that every agent will eventually reach its goal (but not simultaneously with others). 
In terms of performance, \cgamapf outperforms PIBT in dense problems in some cases, while in others, it does not. 
To enjoy the complementary benefits of both algorithms, we propose \cgapibt, a MAPF algorithm that integrates PIBT and \cgamapf into a single, efficient procedure.  %and enjoys the advantages of both approaches. 
% Specifically, \cgapibt enables us to find very efficient single-step moves while maneuvering narrow corridors. [repetition]

% Results show that CGA-MAPF is great
Lastly, we conducted a large set of experiments on standard MAPF benchmarks~\cite{stern2019mapf} comparing \cgamapf and \cgapibt to other standard and state-of-the-art suboptimal algorithms, namely PrP, PIBT, LNS2, LaCAM, and LaCAM$^*$. 
The results show that \cgamapf and \cgapibt can generate outstanding results in terms of success rate in many MAPF benchmarks. For example, in a challenging \emph{maze-32-32-2} grid, the proposed algorithms solved the majority of instances with 450 agents, while the baseline algorithms were not able to solve any of them.

%%%%%%%%%%%%%%%%%%%%%%%%%%%%%%%%%%%%%%%%%%%%
%%%%%%%%%%%%%%%%%%%%%%%%%%%%%%%%%%%%%%%%%%%%
%%%%%%%%%%%%%%%%%%%%%%%%%%%%%%%%%%%%%%%%%%%%
%%%%%%%%%%%%%%%%%%%%%%%%%%%%%%%%%%%%%%%%%%%%
%%%%%%%%%%%%%%%%%%%%%%%%%%%%%%%%%%%%%%%%%%%%
%%%%%%%%%%%%%%%%%%%%%%%%%%%%%%%%%%%%%%%%%%%%
%%%%%%%%%%%%%%%%%%%%%%%%%%%%%%%%%%%%%%%%%%%%
%%%%%%%%%%%%%%%%%%%%%%%%%%%%%%%%%%%%%%%%%%%%
%%%%%%%%%%%%%%%%%%%%%%%%%%%%%%%%%%%%%%%%%%%%
\section{Background}

A \emph{MAPF} problem is defined by a tuple $\langle G, n, s, t \rangle$ where $G \doteq (V, E)$ represents an undirected graph, 
$n$ is the number of agents,
$s: [1, ... , n] \rightarrow V$ maps an agent to a start vertex and $t: [1, ... , n] \rightarrow V$ maps an agent to a target/goal vertex. 
Time is discretized, and in every time step, each agent occupies a single vertex and performs one action. 
% An action is a function $a: V \rightarrow V$ such that $a(v) = v'$, meaning if an agent executes action $a$ from vertex $v$ it will end up in vertex $v'$ in the next time step. 
There are two types of actions: $wait$ and $move$. A $wait$ action means that the agent will stay at the same vertex $v$ at the next time step. A $move$ action means that the agent will move to an adjacent vertex $v'$ in the graph (i.e. $(v, v') \in E$).
A \emph{single-agent plan} for agent $i$, denoted $\pi_i$, is a sequence of actions $\pi_i$ that is applicable starting from $s(i)$ and ending in $t(i)$. 
A solution to a MAPF is a set of single-agent plans $\pi=\{\pi_1,\ldots,\pi_n\}$, one for each agent, that do not have any \emph{conflicts}. 
% Conflicts and a valid solution
We consider two types of conflicts: {\em vertex conflict} and {\em swapping conflict}.
Two single-agent plans have a vertex conflict if they occupy the same vertex at the same time, and a swapping conflict if they traverse the same edge at the same time from opposing directions. 
The objective is to find a solution that minimizes the {\em Sum-of-Costs} (SoC), that is the sum over the lengths of $\pi$'s constituent single-agent paths, or the {\em makespan}, that is the maximum length of all $\pi$'s single-agent paths. 
% Next, we present some common standard MAPF solvers that we used as baselines in this paper.

\paragraph{MAPF algorithms}
Many suboptimal MAPF algorithms have been proposed, including PrP, Large Neighborhood Search (LNS2), LaCAM, and others. 
The algorithm that is mostly related to our work is PIBT~\cite{okumura2022priority}. 
PIBT searches for valid paths in the \emph{configuration} space, where a configuration is a vector representing the agents' locations in some time-step. 
PIBT searches this space in a greedy and myopic manner. 
It starts from the initial configuration of the agents and in every iteration generates a configuration for the next time-step until reaching a configuration in which all agents are at their goals. 
PIBT generates configurations recursively, moving every agent toward its goal while avoiding conflicts with previously planned agents. 
To avoid deadlocks, PIBT utilizes Priority Inheritance and Backtracking techniques. 
% \roni{Maybe explain here that PIBT has the reachability property, and explain what it means}
PIBT is very efficient computationally but is incomplete since it searches greedily in the configuration space. 

\paragraph{The Corridor Generating Algorithm (CGA)}
CGA is an algorithm for solving the Single-Agent Corridor Generating (SACG) problem~\cite{arseni2024corr}. SACG is defined on an environment with multiple agents. The objective is to move a single main agent to its target as fast as possible while avoiding conflicts with all other agents. The other agents have no goals of their own, but they may be required to move to evacuate the path of the main agent. 

CGA is a complete algorithm for solving SAGC~\cite{arseni2024corr}.
% A SACG problem is defined by a tuple $\langle G, s, g, k, \tilde{S} \rangle$, where $G = \langle V, E \rangle$ is a strongly connected undirected graph representing the possible locations the agents may occupy and the allowed transitions between them; $s$ and $g$ are vertices in $G$ representing the start and goal locations of the \emph{main agent}, respectively;  $k$ is the number of agents in addition to the main agent; and $\tilde{S}=(\tilde{S}_1,\ldots,\tilde{S}_k)$ is the vector of vertices in $G$ representing the initial locations of other agents. 
% A solution to a SACG is a set of paths $\pi_0,\ldots,\pi_k$, one per agent, where $\pi_0$ is the path of the main agent and (1) the paths do not conflict, (2) the path of every agent starts from its start location, and (3) the path of the main agent ends in $g$. 
% Paths do not conflict if the agents following them do not occupy the same vertex or edge at the same time. 
% The objective of SACG is to find a solution such that the path of the main agent is minimal.
% Previously, the complete and fast CGA~\cite{arseni2024corr} algorithm was proposed to solve SACG. Our approach builds on this work.
% The \cgamapf algorithm we propose in this work is a rule-based MAPF algorithm that builds on CGA~\cite{arseni2024corr}, an algorithm for solving SACG problems. %Thus, we provide here relevant details on CGA. 
It works by identifying \emph{separating vertices} and \emph{corridors}, which are defined as follows. 

\begin{figure}[!t]
  \centering
  \subfigure[]{\includegraphics[scale=0.16]{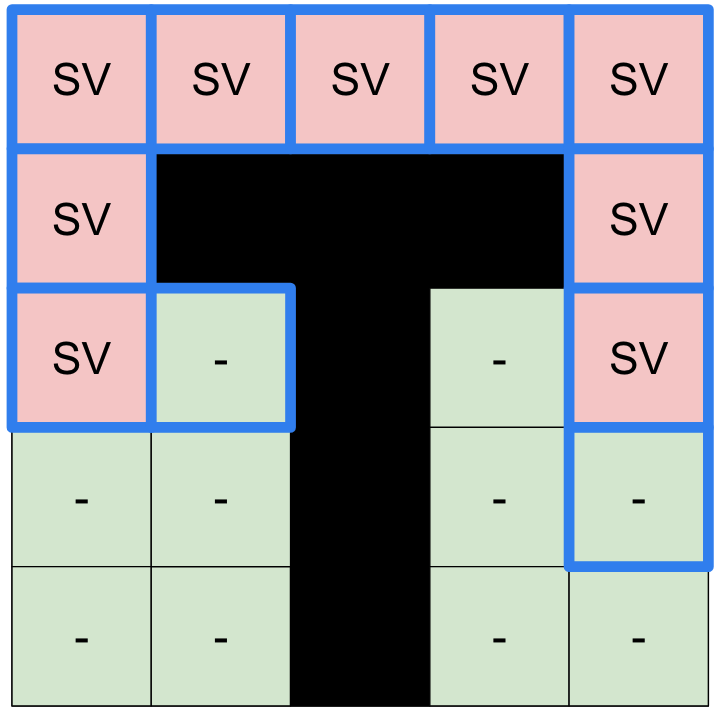}}
  \subfigure[]{\includegraphics[scale=0.16]{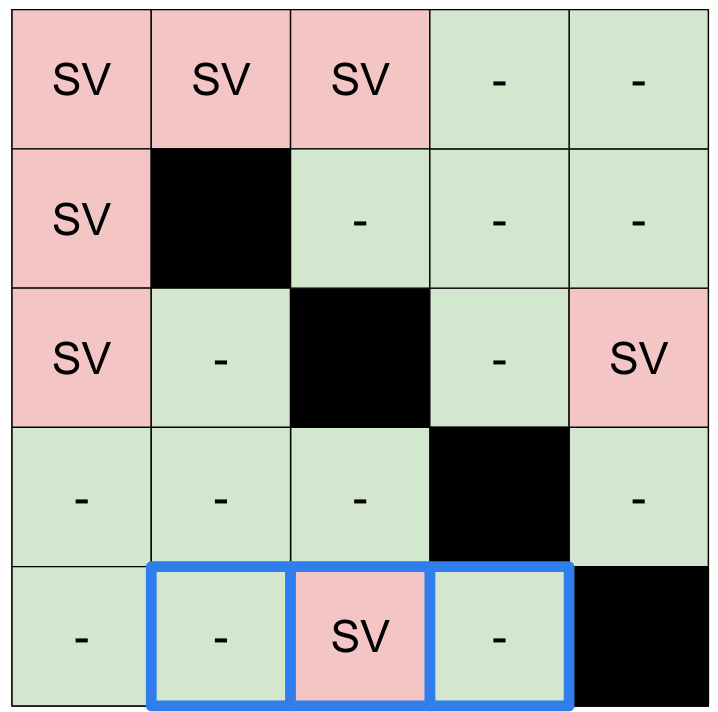}}
  \subfigure[]{\includegraphics[scale=0.16]{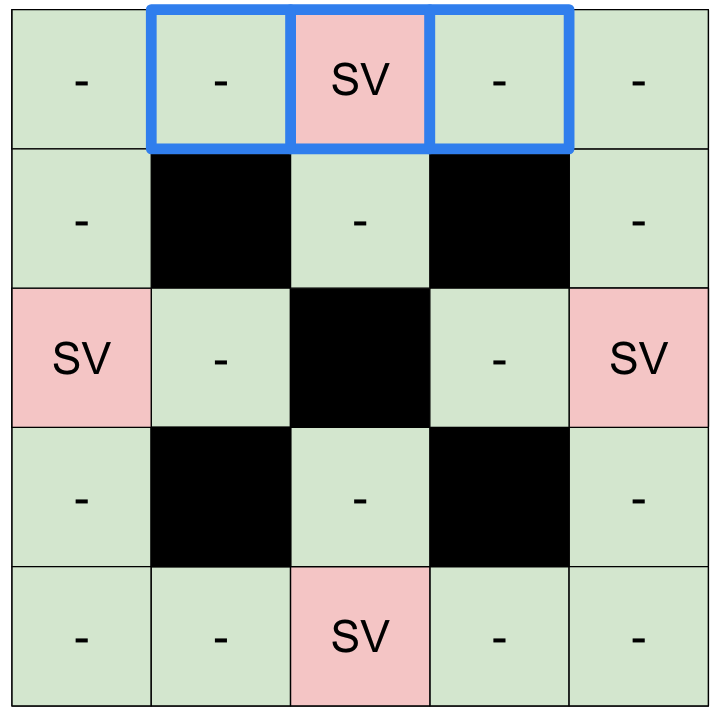}}
  % \subfigure[]{\includegraphics[scale=0.16]{pics/sv3_v2.png}}
  \subfigure[]{\includegraphics[scale=0.175]{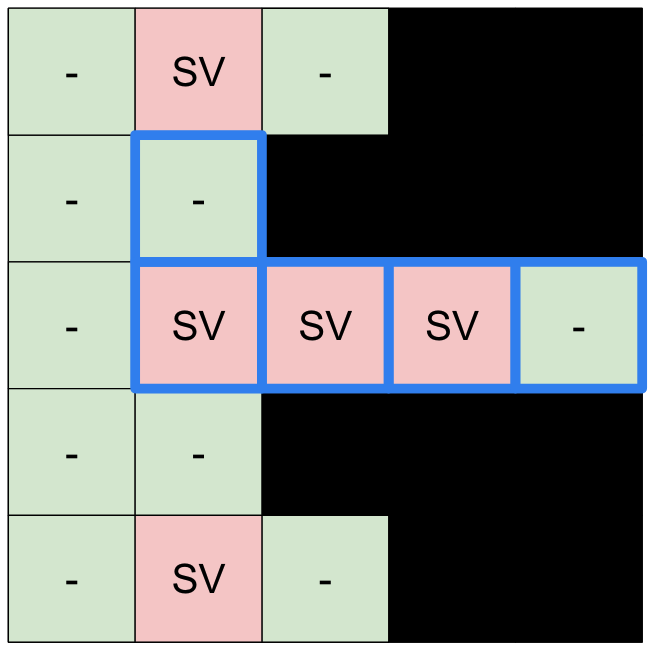}}
  \caption{SVSs for different graphs. Black cells are obstacles; red cells marked by \emph{``SV''} are the SVs; green cells are non-SVs.}
  \label{fig:fm}
\end{figure}

% \begin{definition}[Separating Vertices and Corridors]
% \label{def:sv}
%     A vertex $v$ in a graph $G$ is called a separating vertex (SV) if removing $v$ from $G$ results in a graph with more connected components than $G$.  
%     A corridor in $G$ is a path $(v_1,\ldots,v_n)$ in $G$ such that all the vertices $v_2,\ldots,v_{n-1}$ are SVs. 
%     That is, a corridor is a path in which all vertices except the first and last must be SVs. 
% \end{definition}
\begin{definition}[Separating Vertex]
\label{def:sv}
    A vertex $v$ in a graph $G$ is called a separating vertex (SV) if removing $v$ from $G$ results in a graph $G'$ that includes more connected components than the number of connected components in $G$.
\end{definition}
\begin{definition}[Corridor]
\label{def:corridor}
    A corridor in $G$ is a path $(v_1,\ldots,v_n)$ in $G$ such that all the vertices $v_2,\ldots,v_{n-1}$ are SVs. 
    That is, a corridor is a path in which all vertices except the first and last must be SVs. 
\end{definition}
% \begin{definition}[Trivial Corridor]
% \label{def:trivial_corridor} [this is not such a fundamental concept that deserves a definition]
    A corridor is called \emph{trivial} if it consists of a pair of neighboring vertices $(v_1,v_2)$ in $G$ such that both of the vertices are not SVs. This is the shortest possible corridor, as there are no intermediate SVs between  $v_1$ and $v_2$ 
% \end{definition}
A vertex that is not an SV is denoted as \emph{non-SV} and the set of all SVs for a given graph is denoted as \emph{SVS}.
Fig. \ref{fig:fm} shows examples of SVSs for several grids and highlights in blue some of the corridors in these graphs.  Any subsequent two green cells are an example of a \emph{trivial corridor}.
% \roni{In Figure 1, add a corridor in some way and mention in the text. Then the sentence above would be ``Figure X shows examples of SVSs for several grids and highlights in blue some of the corridors in these graphs}

CGA first finds an optimal path $\pi$ for the main agent to the goal assuming no other agents exist. Path $\pi$ can be decomposed to a sequence of corridors. CGA  iteratively moves the main agent along these corridors one at a time. While moving through a corridor, CGA ensures that the other agents are evacuated from that corridor to allow the main agent to pass through it. The evacuation of other agents can be performed before or in parallel to the moves of the main agent as long as the main agent is guaranteed to be able to pass through the corridor. See Pertzovsky et al.~\shortcite{arseni2024corr} for more details. CGA was proved to be a complete algorithm for solving SACG problems, i.e., it guarantees that the main agent eventually reaches its target, given that the following two conditions hold. The first condition is that the number of non-SVs is larger than the length of the longest corridor in a grid. This condition holds true in most grids that are commonly used in MAPF benchmarks. The second condition is that the initial location of the main agent is a non-SV.

%%%%%%%%%%%%%%%%%%%%%%%%%%%%%%%%%%%%%%%%%%%%
%%%%%%%%%%%%%%%%%%%%%%%%%%%%%%%%%%%%%%%%%%%%
%%%%%%%%%%%%%%%%%%%%%%%%%%%%%%%%%%%%%%%%%%%%
%%%%%%%%%%%%%%%%%%%%%%%%%%%%%%%%%%%%%%%%%%%%
%%%%%%%%%%%%%%%%%%%%%%%%%%%%%%%%%%%%%%%%%%%%
%%%%%%%%%%%%%%%%%%%%%%%%%%%%%%%%%%%%%%%%%%%%
%%%%%%%%%%%%%%%%%%%%%%%%%%%%%%%%%%%%%%%%%%%%
%%%%%%%%%%%%%%%%%%%%%%%%%%%%%%%%%%%%%%%%%%%%
%%%%%%%%%%%%%%%%%%%%%%%%%%%%%%%%%%%%%%%%%%%%
% \section{CORRIDOR GENERATING ALGORITHM FOR MULTI-AGENT PATHFINDING}
\section{Multi-Agent CGA (\cgamapf)}

The {\em Multi-Agent CGA} algorithm solves the classical MAPF problem where all agents need to eventually be at their goals simultaneously. 
It iterates over all agents and applies CGA for each of them. 
\cgamapf works in steps. 
At each step, it associates agents with sequences of one or more actions to perform.  We refer to such a sequence as the \emph{active plan} of an agent. Then, each agent performs the first step of its active plan, and the next step begins.~\footnote {\cgamapf can be activated online where agents physically perform their moves. Alternatively, it can be activated offline, and these moves are added to a future intended plan that is returned at the end of the execution.}

Initially, all agents do not have an active plan. 
Then, \cgamapf alternates between a \emph{planning phase} and an \emph{execution phase}. 
In the planning phase, \cgamapf iterates over all agents. 
For every agent that does not yet have an active plan and attempts to create one for it. 
Creating an active plan for agent $a$ is done as follows. 
First, a \emph{corridor} is identified from the current location of agent $a$' along an optimal path to its goal. 
If that corridor is not occupied, $a$ is assigned an active plan to go through it. 
Otherwise, if the corridor is occupied by agents with no active plan, a dedicated \evacuate procedure is invoked to try to set active plans for these agents who move them out of the corridor. 
If the \evacuate procedure succeeds, \cgamapf assigns $a$ with an active plan to go through the chosen corridor. 
Otherwise, agent $a$ and the agents that block its corridor are not assigned an active plan.  
At the end of the planning phase, agents without an active plan are assigned a default active plan of staying in their place for one time step. 
In the execution phase, all agents perform a single step from their active plans and update their active plans accordingly. 
\cgamapf halts when all the agents reach their goals.
Next, we describe in more detail the key components of the planning phase, namely (1) how to choose a corridor for a given agent $a$, (2) how to try to evacuate the other agents from the corridor, and (3) how to assign $a$ with an active plan that passes through the corridor afterward. 
A detailed pseudo-code of \cgamapf is presented further in this paper.

\begin{figure*} [!t]
 \begin{center}
     \subfigure[Initial state]{\includegraphics[scale=0.4]{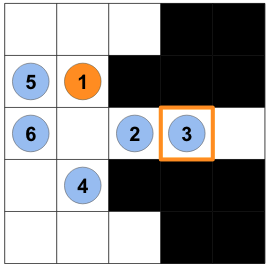}}
     \subfigure[Corridor]{\includegraphics[scale=0.4]{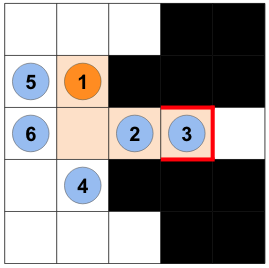}}
     % \subfigure[BFS]{\includegraphics[scale=0.35]{pics/q6.png}}
     \subfigure[EP1]{\includegraphics[scale=0.4]{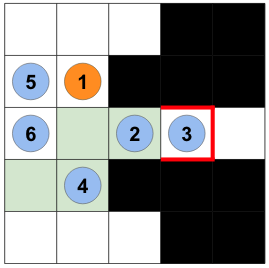}}
     % \subfigure[BFS]{\includegraphics[scale=0.35]{pics/q8.png}}
     \subfigure[EP2]{\includegraphics[scale=0.4]{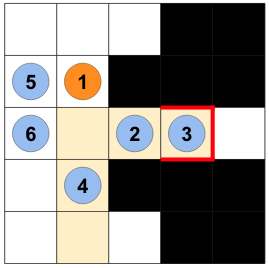}}
     \subfigure[Evac. 1]{\includegraphics[scale=0.4]{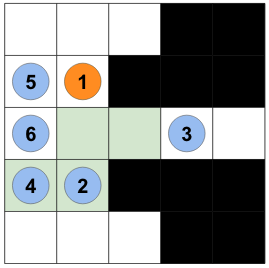}}
     \subfigure[Evac. 2]{\includegraphics[scale=0.4]{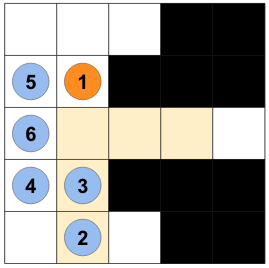}}
     \subfigure[Push]{\includegraphics[scale=0.4]{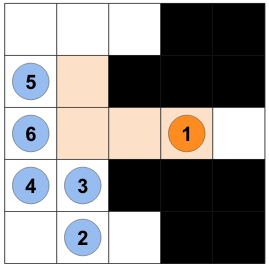}}
     \subfigure[Final state]{\includegraphics[scale=0.164]{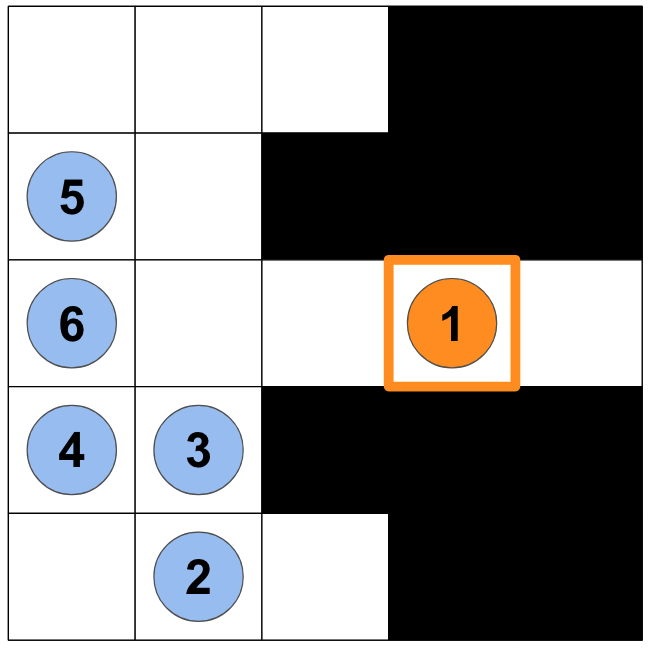}}
 \caption{An example of \cgamapf execution.} 
 % (a) Agent 1 starts from the new initial state.
 % (b) Agent 1 constructs a corridor to its goal. The red lines block the \evacuate from finding EPs that go through the target vertex of agent 1. 
 % corridor.
 % (c)-(d) This time \cgamapf succeeds to find EPs.
 % (e)-(f) \cgamapf evacuates the corridor within those EPs.
 % (g) Agent 1 successfully arrives at its initial target vertex. 
 % % (f)-(i) This time \cgamapf succeeds to find EPs.
 % % (j)-(k) \cgamapf evacuates the corridor within those EPs.
 % % (l) Agent 1 successfully arrives at its initial target vertex.
 % }
 \label{fig:cga_instance_simple}
 \end{center}
 \end{figure*}

\paragraph{Choosing a Corridor} 
This procedure receives agent $a$'s current location and its goal, denoted $a.curr$ and  $a.goal$, respectively.
The procedure, denoted as \createcorridor, outputs a corridor $a.corridor$ that starts from $a.curr$ and ends in either $a.goal$ or a non-SV (as done in CGA). 
Specifically, it chooses an optimal path from $a.curr$ to $a.goal$ (ignoring all other agents). 
Then, it iterates over the vertices in this path starting from $a.curr$ and appending them to $a.corridor$ until reaching either $a.goal$ or a non-SV. 
To identify if a vertex is an SV or non-SV, we compute the SVS for every agent as described by \cite{arseni2024corr} in a pre-processing stage while considering the agent's goal vertex as an obstacle. 
This can be done in polynomial time. 
Fig.~\ref{fig:cga_instance_simple}.a shows an example where the agent needs to go from the orange circle to the orange square. Fig.~\ref{fig:cga_instance_simple}.b shows a corridor for agent 1.

\paragraph{Evacuating the Corridor}
This procedure, referred to as \evacuate, receives an agent $a$; its corridor, $a.corridor$; the locations of the other agents; and their active plans. 
\evacuate attempts to find an active plan for the agents currently occupying $a.corridor$ such that they eventually exit the corridor, allowing $a$ to pass through it. 
If \evacuate returns $False$, it means that it has not been able to free $a.corridor$ from other agents, and $a$ should not be assigned an active plan that passes through it at this point. 

% Finding EPs
Let $A_{in}$ denote the set of agents whose current location resides in $a.corridor$. 
If $A_{in}$ includes an agent that already has an active plan, \evacuate immediately returns $False$. 
Otherwise, it iterates over the agents in $A_{in}$, and for each agent $a'\in A_{in}$, it searches for a path from its current location to an unoccupied vertex outside of $a.corridor$.
To avoid deadlocks and livelocks, this search is not allowed to pass through any vertex planned to be occupied by an active plan of another agent or to pass through edges connected to $a$'s target vertex that lead outside of $a.corridor$.
% does not pass through the target of agent $a$ or any vertex planned to be occupied by an active plan of another agent. 
We refer to such a path as an \emph{evacuation path} (EP). 
Importantly, \evacuate ensures that each EP ends at a different vertex. 
Finding EPs that satisfy these requirements is done by running a Breadth-First Search for each agent from its current location, ignoring conflicts with other agents in $A_{in}$, and avoiding the prohibited vertices and edges mentioned above. 
If the search for an EP fails for some agent $a'\in A_{in}$, \evacuate returns $False$. 
Otherwise, we obtain a set of EPs, one for every agent in $A_{in}$. 

% Why we are to done
Note that an EP is a path in $G$, while an active plan is a sequence of actions that may include wait and move actions. 
Based on the EPs found, \evacuate creates an active plan for every agent $a'$ in $A_{in}$, denoted $\pi(a')$, as follows. 
Initially, these active plans are empty. 
Then, \evacuate iterates over the EPs in some arbitrary order, 
performing the following steps for every EP $e$. 

For each agent $a' \neq a$, we use $loc(a')$ to denote the last location that $a'$ is expected to reach after following its active plan. 
We iterate through the vertices of $e$ from the closest to the current location of $a$ to the furthest away from it, and for every vertex $e_i$, we check if there is an agent $a'$ for which $loc(a')$ equals $e_i$.
If such an agent exists, we add it to an ordered list $e(A)$, which is sorted according to the distance of $loc(a')$ from the current location of the agent $a$.
Once the procedure is completed and all relevant agents have been added to $e(A)$ (assume that there are $m$ agents in $e(A)$ and denote the last among them by $a_m$)
% For an EP $e$, \evacuate identifies agents $a'$ for which $loc(a')$ ends up inside $e$. 
% $e(A)\doteq (a_1,\ldots, a_n)$ is the list of these agents, sorted by order of $loc(\cdot)$ in $e$. 
we append to $\pi(a_m)$ a sequence of actions moving $a_m$ from $loc(a_m)$ to the last vertex in $e$. 
Following this procedure ensures that the vertex $loc(a_m)$ is now unoccupied by any agent. 
Next, we iterate over the other agents in $e(A)$ in reverse order, appending to the active plan of agent $a_i$ a sequence of actions that move it from $loc(a_i)$ to the previous location of $a_{i+1}$. 
Whenever we append an action to an active plan of an agent $a_i$, we check if an active plan of another agent intends to occupy the same vertex in the future. This may occur when EPs overlap. 
In such a case, we append wait actions to the active plan of $a_i$ until it can safely move to that vertex. 
% By definition, $loc(a_1)$ is the first vertex in $e$. 
Note that the above process is repeated for every EP, potentially appending actions to agents in $A_{in}$ multiple times. 
Fig.~\ref{fig:cga_instance_simple}(c-d) shows examples of finding EPs and 
Fig. \ref{fig:cga_instance_simple}(e-f) shows the execution of the corresponding active plans.

\paragraph{Passing through the Corridor}
In the case where \evacuate was able to successfully create active plans, the next step involves assigning the agent $a$ with an active plan that moves it through the chosen corridor. Similar to how actions are added to agents in \evacuate, we only add an action to $a$'s active plan to move to a vertex $v$ if $v$ is not occupied in the future by any active plan. Otherwise, we append a wait action to $a$'s active plan until $v$ is free. 
We denote this procedure as \push.

\paragraph{Temporary Targets} 
As described above, \evacuate may fail (return $False$) to evacuate a corridor $a.corridor$ if it cannot find an EP for some agent $a'$ who resides in the corridor. 
If this failure is not due to an active plan of some other agent, 
there is a risk of reaching a deadlock situation. 
To mitigate this to some extent, if the search for an EP fails but the corresponding search did not attempt to generate any vertex used by an active plan, then  \evacuate also assigns a new, temporary target for $a$ to avoid deadlocks. 
In our implementation, the temporary target was the unoccupied non-SV, which is closest to $a$. 
When an agent reaches a temporary target, it is reassigned its original target. 

\begin{figure} [b!t]
 \begin{center}
     \subfigure[Initial state]{\includegraphics[scale=0.4]{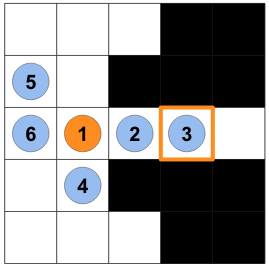}}
     \subfigure[Corridor]{\includegraphics[scale=0.4]{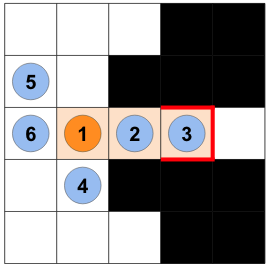}}
     \subfigure[Temp. goal]{\includegraphics[scale=0.4]{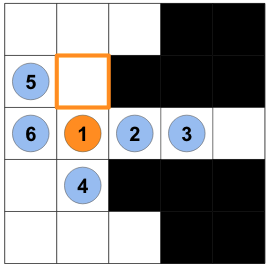}}
     \subfigure[New state]{\includegraphics[scale=0.4]{pics/q4.png}}
 \caption{An example of assigning a temporary goal.}
 % \cgamapf execution. 
 % (a) An initial problem. An orange circle is the highest-order agent with an index of 1. An orange square is the target vertex of the agent. The other agents are presented as blue circles with their indices written inside them. 
 % (b) Agent 1 constructs a corridor to its goal. The red lines block the \evacuate from finding EPs that go through the target vertex of agent 1.
 % (c) \cgamapf cannot evacuate this corridor, so another temporary goal is chosen for agent 1. 
 % (d) Agent 1 starts from the new initial state.
 % (e) Agent 1 constructs a new corridor.
 % (f)-(i) This time \cgamapf succeeds to find EPs.
 % (j)-(k) \cgamapf evacuates the corridor within those EPs.
 % (l) Agent 1 successfully arrives at its initial target vertex.
 % }
 \label{fig:cga_instance_temp_goal}
 \end{center}
 \end{figure}

An example of such an unsolvable instance and how a temporary goal helps to resolve it is depicted in Fig. \ref{fig:cga_instance_temp_goal}. 
Here, agent $1$ has a goal vertex (orange square), but it is impossible to evacuate agents $2$ and $3$ out of the corridor, so the agent moves to the temporary goal vertex (Fig. \ref{fig:cga_instance_temp_goal} (a)-(d)). 
At this stage, we reach the same state as depicted in Fig.~\ref{fig:cga_instance_simple}, which \cgamapf can solve.

\paragraph{Pseudocode} 
\begin{algorithm}[ht]
    \caption{\cgamapf\textcolor{blue}{+PIBT}}
    % \textbf{Input}: $\langle F_{D_i}, F_C \rangle$\\
    % \textbf{Parameter}: Optional list of parameters\\
    % \textbf{Output}: $\pi_i$
    \begin{algorithmic}[1] %[1] enables line numbers
        \STATE \textbf{Input}: $\langle A, G := (V, E) \rangle$\label{line:input}
        \STATE $SVS \leftarrow CreateSVS(G)$ \label{line:svs}
        \WHILE{not all agents at their goals}\label{line:while}
            % \STATE $Q^{new} \leftarrow$ update with agents that have a plan
            \STATE $i \leftarrow$ current time step \label{line:iter}
            \FOR{every $a \in A$}\label{line:qnew} \label{line:for1}
                % \IF{$Q^{new}(a) \neq \emptyset$}\label{line:if1}
                \IF{$a.path[i] \neq \emptyset$}\label{line:if1}
                \STATE \textbf{Continue}
                \ENDIF \label{line:if1ends}
                \IF{$a.curr = a.goal \land a.tempGoal$} \label{line:if2}
                    \STATE $a.tempGoal \leftarrow False; \: a.goal \leftarrow a.initGoal$
                \ENDIF \label{line:if2ends}
                \textcolor{blue}{
                \IF{next node is \emph{non-SV}} \label{line:ifsv_start}
                    \STATE {\tt PIBT($a$, $A$)} \label{line:pibt_step}
                    \STATE \textbf{Continue}
                \ENDIF \label{line:ifsv_end}
                }
                % \IF{next node is \emph{SV}} \label{line:ifsv_start}
                \STATE {\tt CreateCorridor($a$)} \label{line:createcorridor}
                \STATE {\tt found\_bool $\leftarrow$ Evacuate($a$, $A$)}
                \IF{$\neg$ {\tt found\_bool}} \label{line:if3}
                    \STATE \textbf{Continue}
                \ENDIF \label{line:if3ends}
                \STATE {\tt Push($a$, $A$)} \label{line:evacuateandpush}
                % \ELSE
                %     \STATE {\tt PIBT($a$, $A$)} \label{line:pibt_step}
                % \ENDIF \label{line:ifsv_end}
            \ENDFOR \label{line:for1ends}
            \STATE {\tt Execute($i$, $A$)} \label{line:execute}
            \STATE {\tt UpdateOrder($A$) } \label{line:updateorder}
        \ENDWHILE \label{line:whileends}
        \STATE \textbf{Return} $\pi$ \label{line:return}
    \end{algorithmic}
    \label{alg:cga-mapf}
\end{algorithm}

The high-level pseudocode of \cgamapf (excluding the blue text) and \cgapibt (including the blue text) is illustrated in Algorithm \ref{alg:cga-mapf}.
% \roni{The reader does not know yet what is CGAPIBT}
The algorithm starts by creating SVS (line \ref{line:svs}). 
% or uploads it if saved previously .
The algorithm halts only when all the agents are at their goal locations (line \ref{line:while}). 
In every time step, it loops through all agents (lines \ref{line:for1}-\ref{line:for1ends}).
% $Q^{new}$ tracks all the agents that already have a plan for a given time step (line \ref{line:qnew}).
% For each agent $a$ (lines \ref{line:for1}-\ref{line:for1ends}), 
Algorithm~\ref{alg:cga-mapf} continues to the next agent if agent $a$ (line \ref{line:for1}) already has an active plan (lines \ref{line:if1}-\ref{line:if1ends}).
If $a$ is at its goal but the goal is temporary, $a$ sets back its goal to be the initial one (lines \ref{line:if2}-\ref{line:if2ends}).
In \cgapibt, if the next vertex is non-SV, the algorithm executes PIBT on agent $a$ and continues to the next agent (lines \ref{line:ifsv_start}-\ref{line:ifsv_end}).
Then, \cgamapf creates a corridor for a by the \createcorridor procedure (line \ref{line:createcorridor}) and attempts to evacuate the corridor with \evacuate procedure.
If \evacuate fails, the algorithm continues to the next agent (lines \ref{line:if3}-\ref{line:if3ends}).
After that, the \push procedure is executed (line \ref{line:evacuateandpush}).
Lastly, all agents execute their plan for time-step $i$ (the agents without an active plan remain in their current locations)(line \ref{line:execute}). 
The {\tt UpdateOrder} function sends all finished agents to the end of the order (line \ref{line:updateorder}).
Due to the reachability property, a first agent in order is guaranteed to eventually reach its goal.
The {\tt UpdateOrder} ensures that every agent at some point will be first in the order.
If succeeds, the algorithm returns a set of paths for every agent (line \ref{line:return}).

% \paragraph{THEORETICAL PROPERTIES}
\section{Theoretical Properties}
\label{section:theor-prop}

% RUNTIME
First, we analyze the runtime of the \cgamapf procedures. 
The runtime complexity of the \createcorridor procedure is $O(|V| + |E|)$ as it simply runs a Breadth-First Search. 
Similarly, running \evacuate for a single agent requires $O(|V| + |E|)$. \evacuate  searches for EVs at most $|A|$ times, and thus its runtime is $O(|A|(|V|+|E|))$. 
The runtime of \push is at most $O(|A||V|)$, pushing the agents across already calculated corridors.
Unfortunately, there are no guarantees on the global runtime of \cgamapf, as it cannot identify the unsolvable instances. 
% The example of such an instance is illustrated in Figure \ref{fig:unsolv}. 
% Here, \cgamapf will run the search indefinitely. 
% \begin{figure}[!ht]
%   \centering
%   \includegraphics[scale=0.2]{pics/unsolv.png}
%   \caption{An unsolvable MAPF problem. The arrows point to the goal locations of the agents. \cgamapf will not identify the instance as unsolvable.}
%   \label{fig:unsolv}
% \end{figure}

While \cgamapf is incomplete, it does satisfy the \emph{reachability} property as PIBT. 
That is, each agent is guaranteed to eventually reach its goal, albeit possibly not at the same time. 
% The proof is given in the supplementary material. 

% CGA-MAPF has a reachability property
Next, we gradually prove the reachability of \cgamapf. 
\begin{lemma}
\label{lem:compl}
In \cgamapf, if the first agent in order $a_1$ is occupying a non-SV and the number of unoccupied vertices in a graph $G$ is greater than or equal to the length of the longest corridor in $G$, then the \emph{FindEVs} procedure will successfully find EVs for all agents from any corridor for agent $a_1$. 
\end{lemma}
\setlength\parindent{0pt} \textbf{Proof outline.} 
Since the main agent is not occupying a SV, there exists a path from every vertex in the next corridor to any vertex in $G$ that does not go through the main agent's location. As there are more unoccupied vertices than vertices in the corridor, there exists an unoccupied vertex in $G$ for every vertex in this corridor. Thus, \emph{FindEVs} will find evacuation routes for every vertex in the corridor, as required. 
$\square$

\begin{theorem}[Completeness for $a_1$]
\label{theorem:compl-cga}
If the first agent in order $a_1$ is not occupying an SV and the number of unoccupied vertices in a graph is equal to or greater than the length of the longest corridor in $G$, then \cgamapf is guaranteed to bring $a_1$ agent to its goal.
\end{theorem}
{\bf Proof outline.} 
The \emph{CretateCorridor} procedure in CGA ensures that the main agent moves from one non-SV vertex to another along an optimal path to the goal. 
Due to Lemma~\ref{lem:compl}, \emph{FindEVs} together with \emph{EvacuateAndPush} will successfully evacuate the corridor connecting these two non-SV vertices. Consequently, after a finite number of steps, 
the $a_1$ agent will reach its goal. 
$\square$

\begin{theorem}[Reachability of \cgamapf]
In \cgamapf, if the number of unoccupied vertices is larger than the longest corridor and the \emph{UpdateOrder} function ensures that every agent will eventually be first in order, then every agent is guaranteed to reach its next goal location in a finite amount of time.
\end{theorem}
{\bf Proof:} 
Following Theorem~\ref{theorem:compl-cga}, the agent with the highest priority will reach its goal location in a finite amount of steps, as it applies \cgamapf without any restrictions. 
\emph{UpdateOrder} function assigns the lowest priority to agents that have reached their goals. 
Thus, eventually, every agent will be the highest priority agent and reach its goal~\footnote{Note that, in contrast to the completeness property, reachability does not guarantee that the agents will reach their goals simultaneously.}. 
$\square$

% \begin{theorem}[Reachability of \cgamapf]
% In \cgamapf, if the number of unoccupied vertices is larger than the longest corridor and the \emph{UpdateOrder} function ensures that every agent will eventually be first in order, then every agent is guaranteed to reach its next goal location in a finite amount of time.
% \end{theorem}

% \paragraph{COMPARISON WITH MAPF SOLVERS}
\paragraph{Discussion}
\cgamapf can be viewed as a rule-based MAPF algorithm such as Push \& Swap (PS)~\cite{luna2011push} and Push \& Rotate (PR)~\cite{de2013push}, since its \evacuate procedure is somewhat reminiscent of their ``swapping'' and ``rotating'' procedures. 
A key advantage of \cgamapf over these rule-based MAPF solvers is that it pushes all agents to act concurrently as much as possible, mitigating to some extent the poor solution quality often exhibited by rule-based MAPF solvers. 
\cgamapf also bears similarity to prioritized planning algorithms, since 
in its planning phase the agents plan sequentially, blocking future agents from occupying some locations. Unlike PrP, in \cgamapf agents are not required to create full paths to their targets, reducing computational effort. 

\cgamapf have some similarities to PIBT~\cite{okumura2022priority}, as it plans for every agent only a few steps ahead. However, in PIBT the planning is only a single step ahead, which leads to deadlocks and livelocks in MAPF, especially in narrow closed corridors. \cgamapf delicately solves these cases with its \evacuate procedure. 
However, \cgamapf orders agents arbitrarily while PIBT uses heuristics to prioritize agents' movements. This is very beneficial when we are not close to a long corridor.  
To enjoy the complementary benefits of PIBT and \cgamapf, we developed a hybrid algorithm called \cgapibt. 
In \cgapibt, when planning for an agent $a$, we first use PIBT to choose the next vertex to go to, considering the vertices of the other active plans as obstacles. If the vertex chosen by PIBT is a non-SV, we set the active plan to go to it, forming a trivial corridor of size 2. 
Otherwise, we plan for that agent as in the regular \cgamapf algorithm. 

\section{Experimental Results}

% setting 
% We conducted an experimental evaluation comparing \cgamapf within 
% PrP~\cite{prp_2015}, 
% LNS2~\cite{li2022mapf}, 
% PIBT~\cite{okumura2022priority}, LaCAM~\cite{okumura2023lacam}, and LaCAM$^*$~\cite{Okumura_2023}, where PrP and LNS2 are implemented with SIPPS~\cite{li2022mapf}. 
We conducted an experimental evaluation comparing \cgamapf with PrP, LNS2, PIBT, LaCAM, and LaCAM$^*$, where PrP and LNS2 are implemented with SIPPS~\cite{li2022mapf}.
The chosen baselines are considered to be state-of-the-art algorithms for solving MAPF suboptimality~\cite{stern2019mapf,okumura2023lacam}.
We did not include other rule-based MAPF algorithms such as PS~\cite{luna2011push} and PR~\cite{de2013push}
in our baselines as they were shown to be significantly inferior to PIBT~\cite{okumura2022priority}.

\begin{figure} [t]
 \begin{center}
    \subfigure[]{\includegraphics[scale=0.6]{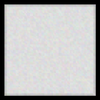}}
    \subfigure[]{\includegraphics[scale=0.6]{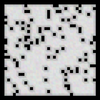}}
    \subfigure[]{\includegraphics[scale=0.6]{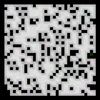}}
    \\
    \subfigure[]{\includegraphics[scale=0.6]{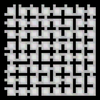}}
    \subfigure[]{\includegraphics[scale=0.6]{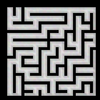}}
    \subfigure[]{\includegraphics[scale=0.6]{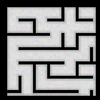}}
 \caption{MAPF Grids: (a) \emph{empty-32-32}, (b) \emph{random-32-32-10}, (c) \emph{random-32-32-20}, (d) \emph{room-32-32-4}, (e) \emph{maze-32-32-2}, (f) \emph{maze-32-32-4}}
 \label{fig:maps}
 \end{center}
 \end{figure}

% Baselines
% Benchmark
All experiments were performed on six different maps from the MAPF benchmark~\cite{stern2019mapf}: \emph{empty-32-32}, \emph{random-32-32-10}, \emph{random-32-32-20}, \emph{room-32-32-4}, \emph{maze-32-32-2}, and \emph{maze-32-32-4} as they present different levels of difficulty.
% \footnote{Due to space limitations, we report the results on \emph{empty-32-32}, \emph{room-32-32-4}, and \emph{maze-32-32-4} grids, and the rest are presented in the supplementary material.}
The maps are visualized in Figure~\ref{fig:maps}. 
The number of agents used in our experiments varied from 100 to 700. 
We executed 20 random instances per every number of agents, map, and algorithm.
A time limit of 30 seconds was imposed on every instance. 
The ratio of instances solved within this time limit by a given algorithm is referred to as the ``success rate'' of that algorithm. 
All algorithms were implemented in Python and ran on a MacBook Air with an Apple M1 chip and 8GB of RAM.\footnote{Prior works implemented MAPF algorithms on different programming languages, including C++, Java, C\#, and Python. Thus, our results cannot be compared blindly with the published results from other papers. Nevertheless, we made significant efforts to verify that our implementations carefully matched published implementations of other algorithms if such existed.}
% and the original code of authors

% \begin{figure}[!ht]
%     \centering
%     \includegraphics[scale=0.0991]{pics/sr_empty.png}
%     \includegraphics[scale=0.0991]{pics/sr_room.png}
%     \includegraphics[scale=0.0991]{pics/sr_maze4.png}
%     \\
%     \includegraphics[scale=0.0991]{pics/rt_empty.png}
%     \includegraphics[scale=0.0991]{pics/rt_room.png}
%     \includegraphics[scale=0.0991]{pics/rt_maze4.png}
%     \\
%     \includegraphics[scale=0.0991]{pics/ms_empty.png}
%     \includegraphics[scale=0.0991]{pics/ms_room.png}
%     \includegraphics[scale=0.0991]{pics/ms_maze4.png}    
%   \caption{Success rate (top), runtime (middle), and makespan (bottom) results.}
%   \label{fig:sr-rt-ms}
% \end{figure}

% % success rate
\begin{figure}[!t]
    \centering
    \includegraphics[scale=\graphsize]{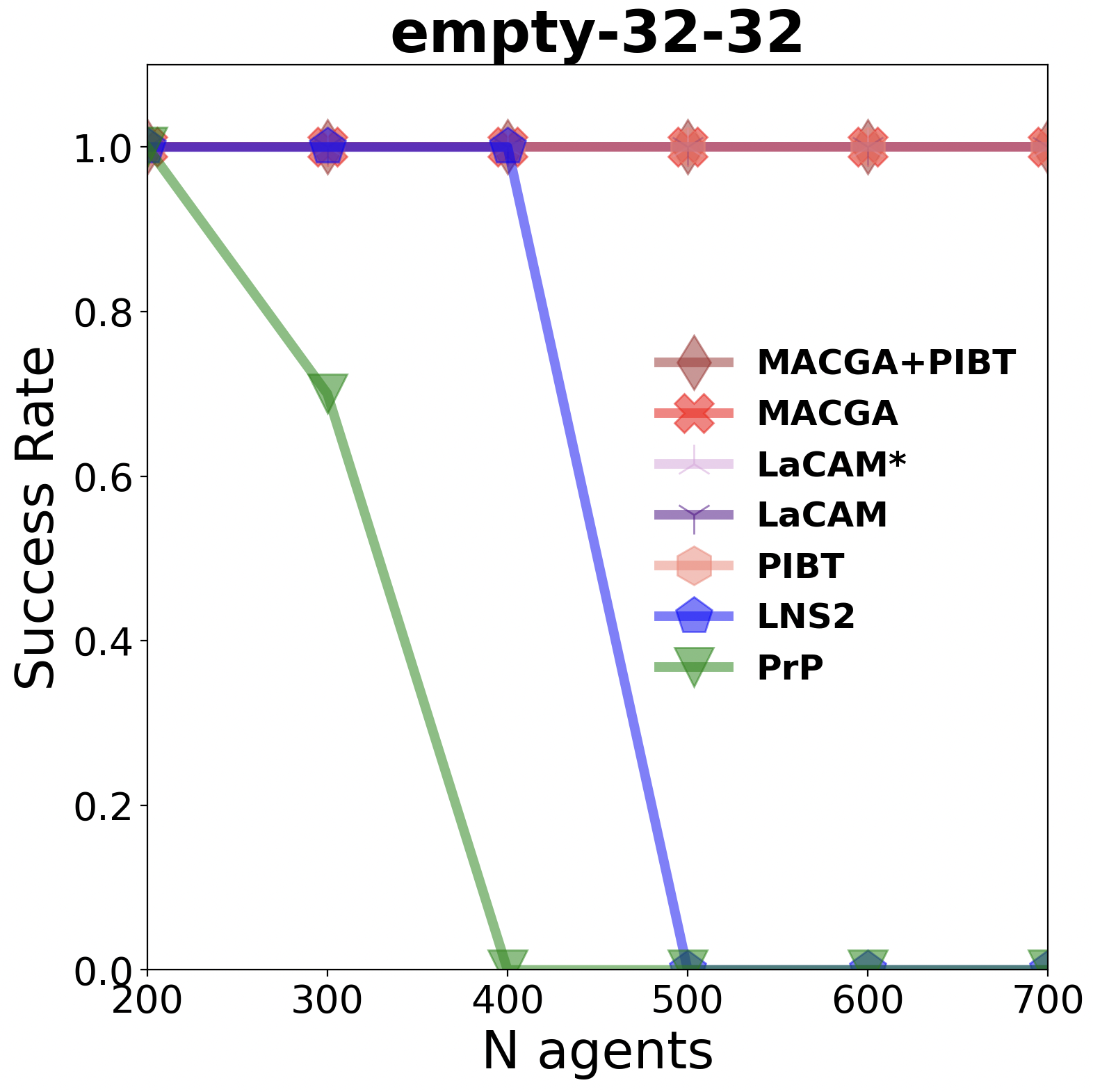}
    \includegraphics[scale=\graphsize]{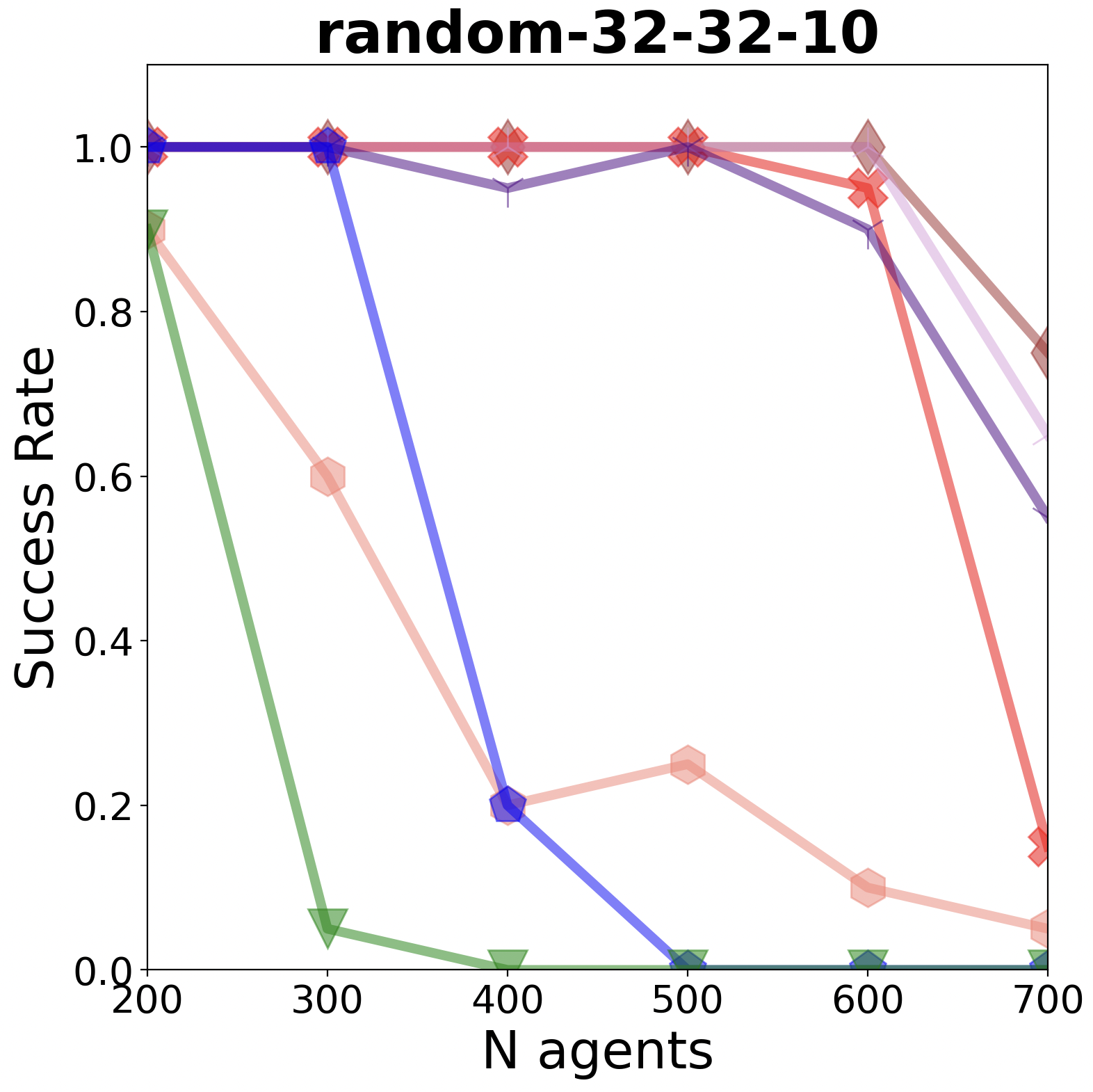}
    \includegraphics[scale=\graphsize]{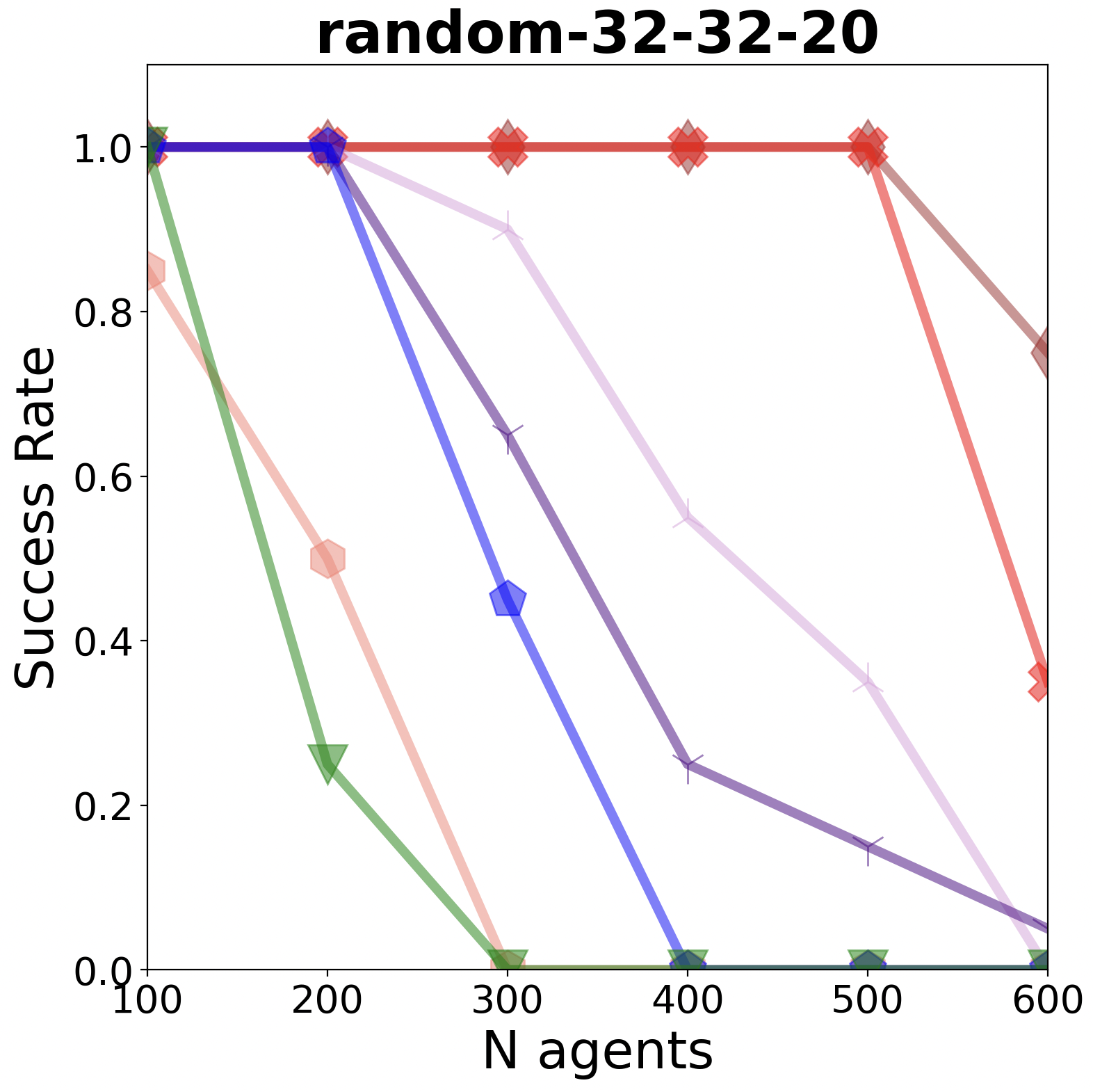}
    \includegraphics[scale=\graphsize]{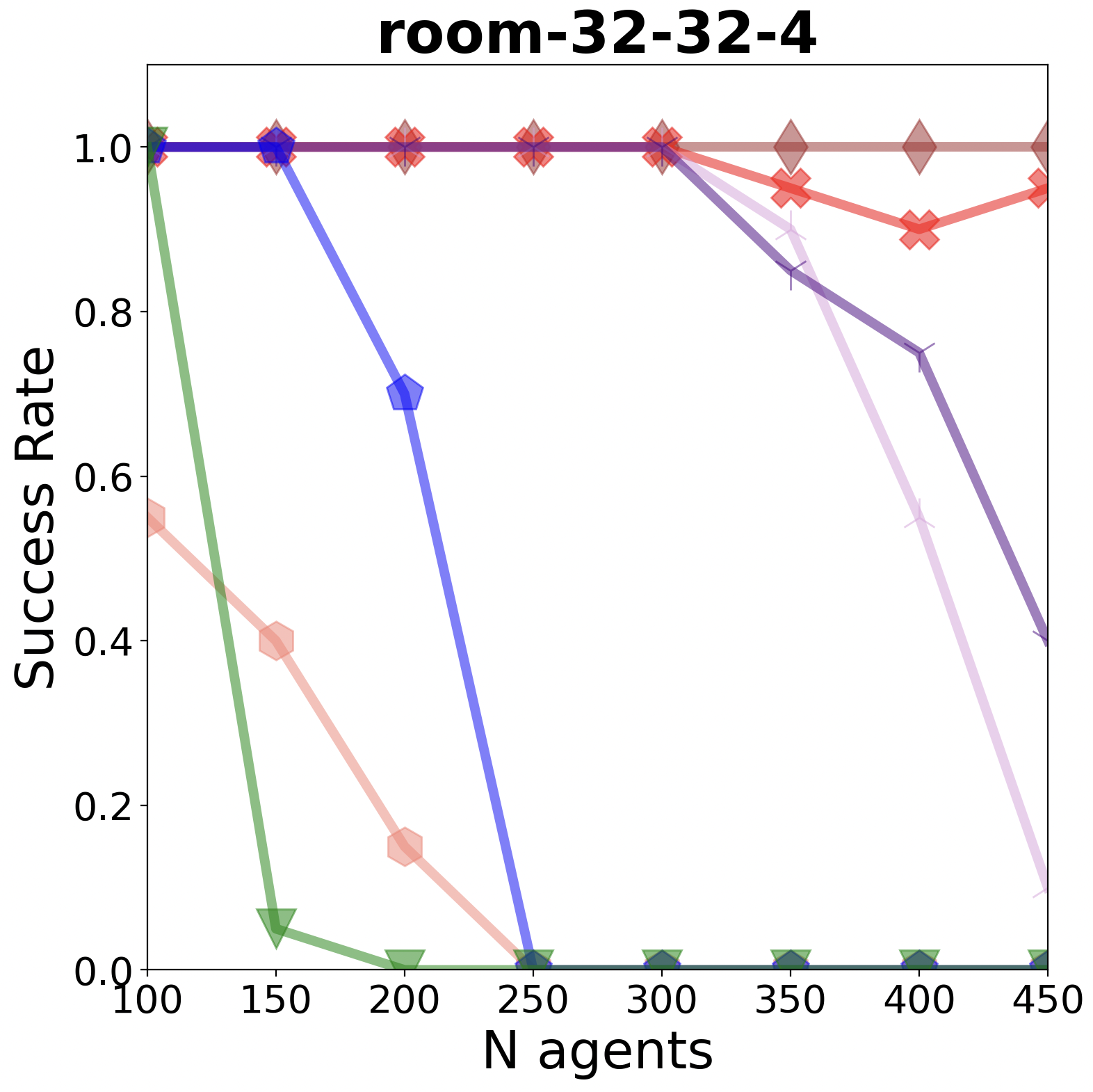}
    \includegraphics[scale=\graphsize]{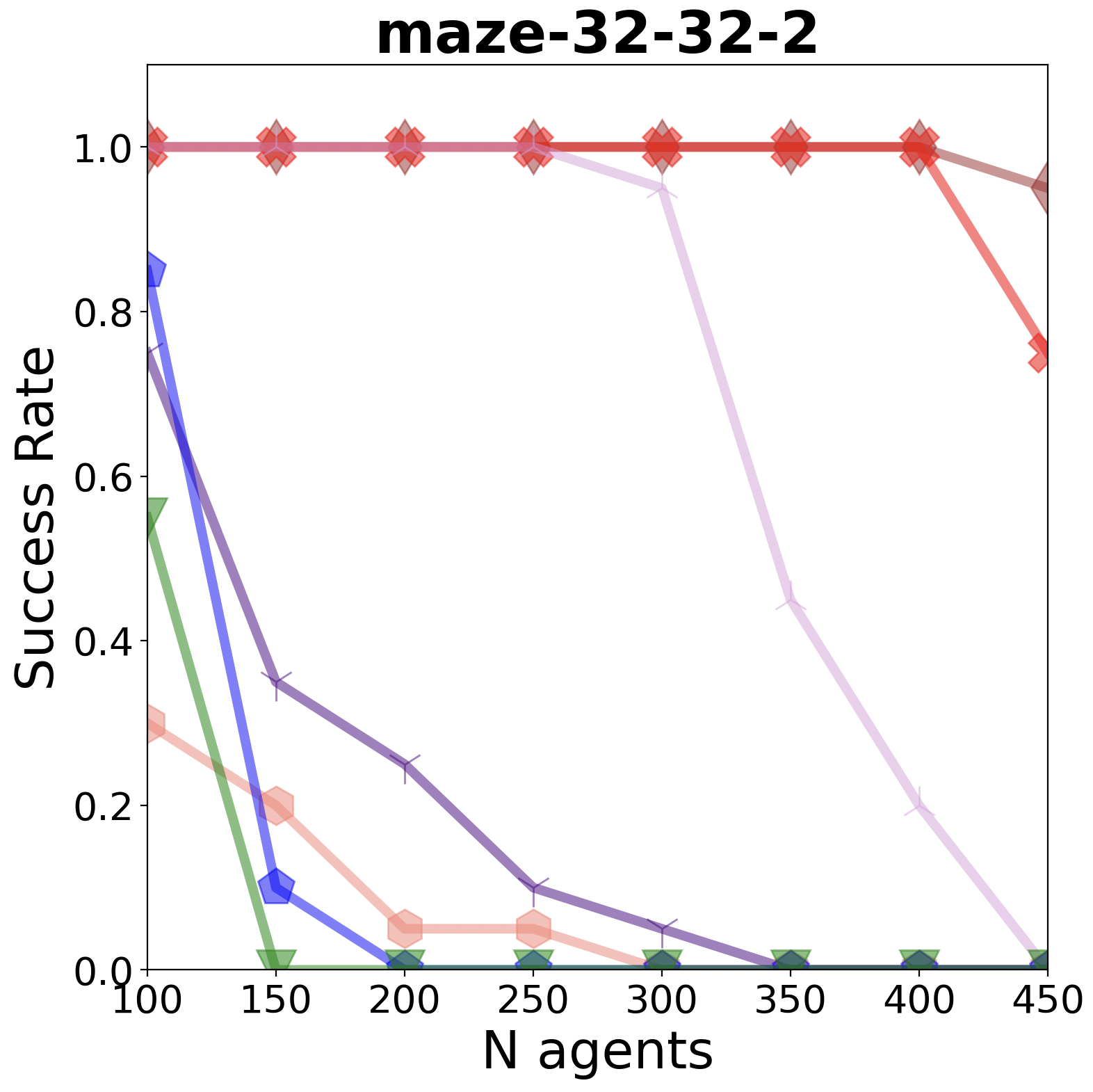}
    \includegraphics[scale=\graphsize]{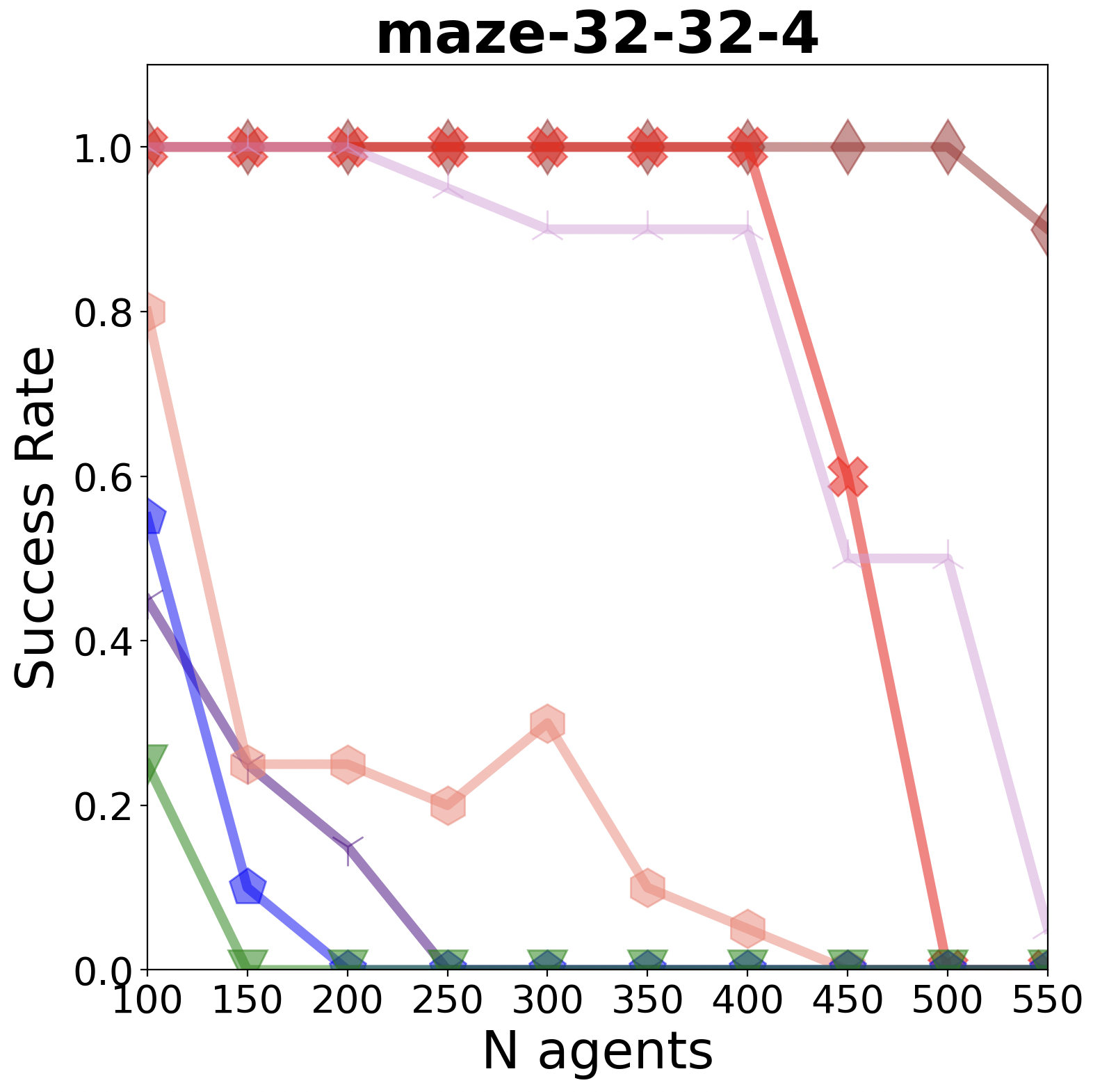}
  \caption{Success Rate}
  \label{fig:sr}
\end{figure}
Fig.~\ref{fig:sr} presents the {\em success rate} (SR) of algorithms ($y$-axis) in different grids per number of agents ($x$-axis), where the SR is the ratio of problems that could be solved within the allocated time limit.  
In general, the results show that  \cgamapf and \cgapibt solve the majority of the problems, outperforming all the others evaluated algorithms.
The best performance of our approaches relative to baseline algorithms was in the grids that contain narrow spaces, such as \emph{room-32-32-4} or \emph{maze-32-32-2}.
As an example, in \emph{maze-32-32-2} grids, our approaches solved most of the instances within 450 agents, while other baseline algorithms did not succeed in solving any of them.
\cgapibt outperforms \cgamapf in terms of success rate in almost all maps. 
This suggests the goal-oriented behavior of PIBT in trivial corridors indeed yields the intended effect of moving the agents faster toward their goal.
% This can be explained by the fact that \cgapibt utilizes a more efficient PIBT algorithm that is executed for \emph{trivial} corridors.

% % runtime
\begin{figure}[!t]
    \centering
    \includegraphics[scale=\graphsize]{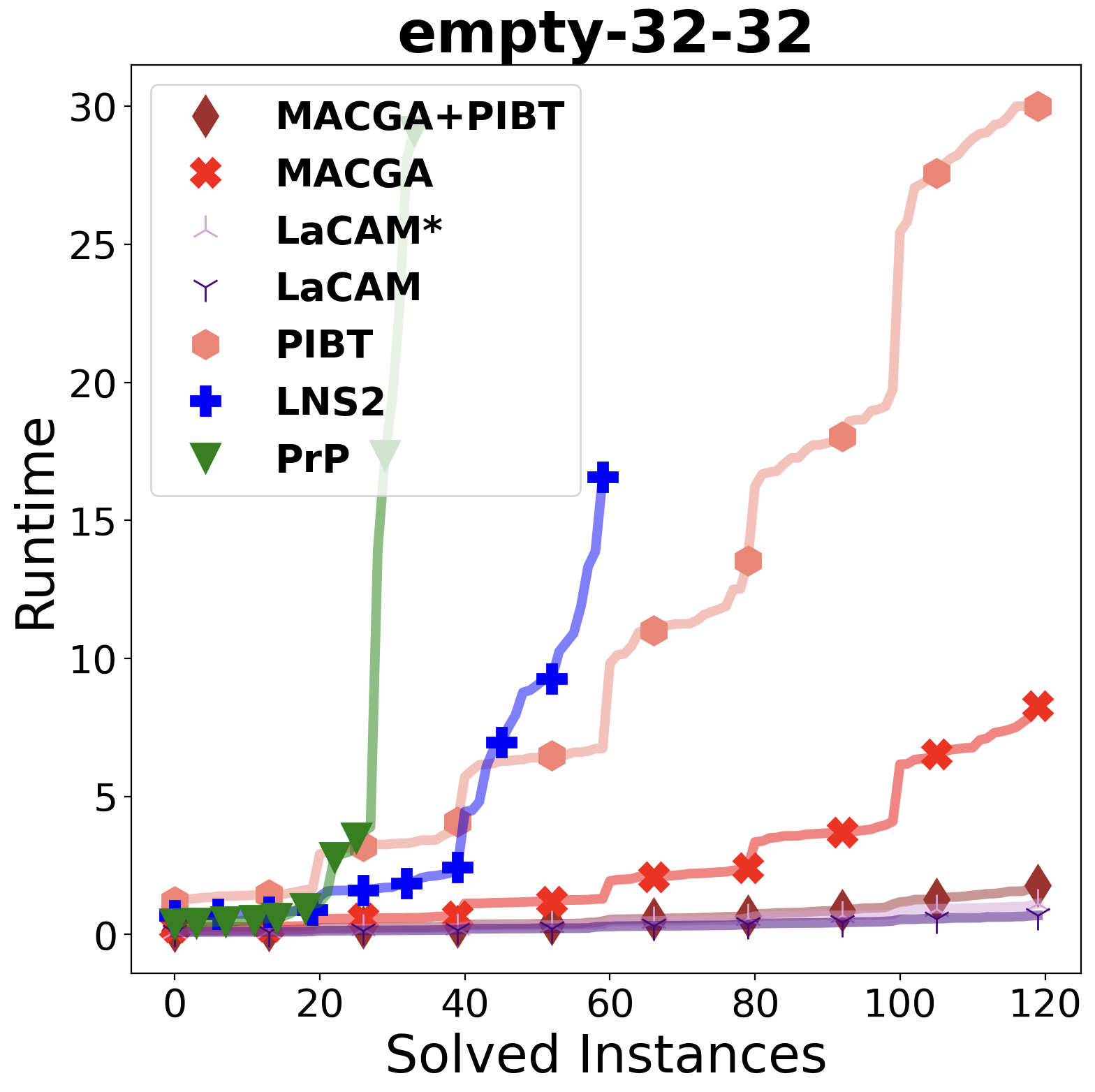}
    \includegraphics[scale=\graphsize]{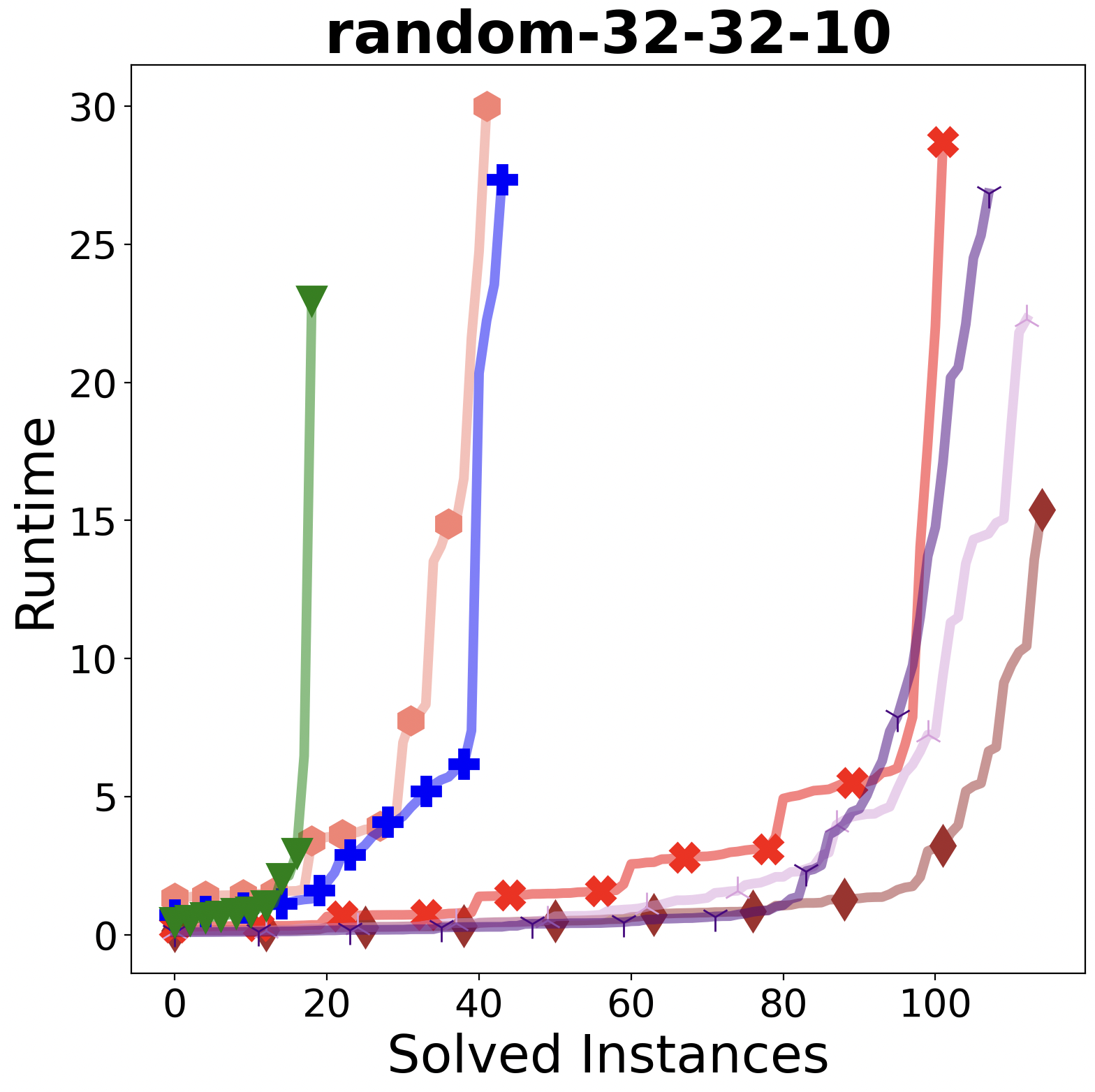}
    \includegraphics[scale=\graphsize]{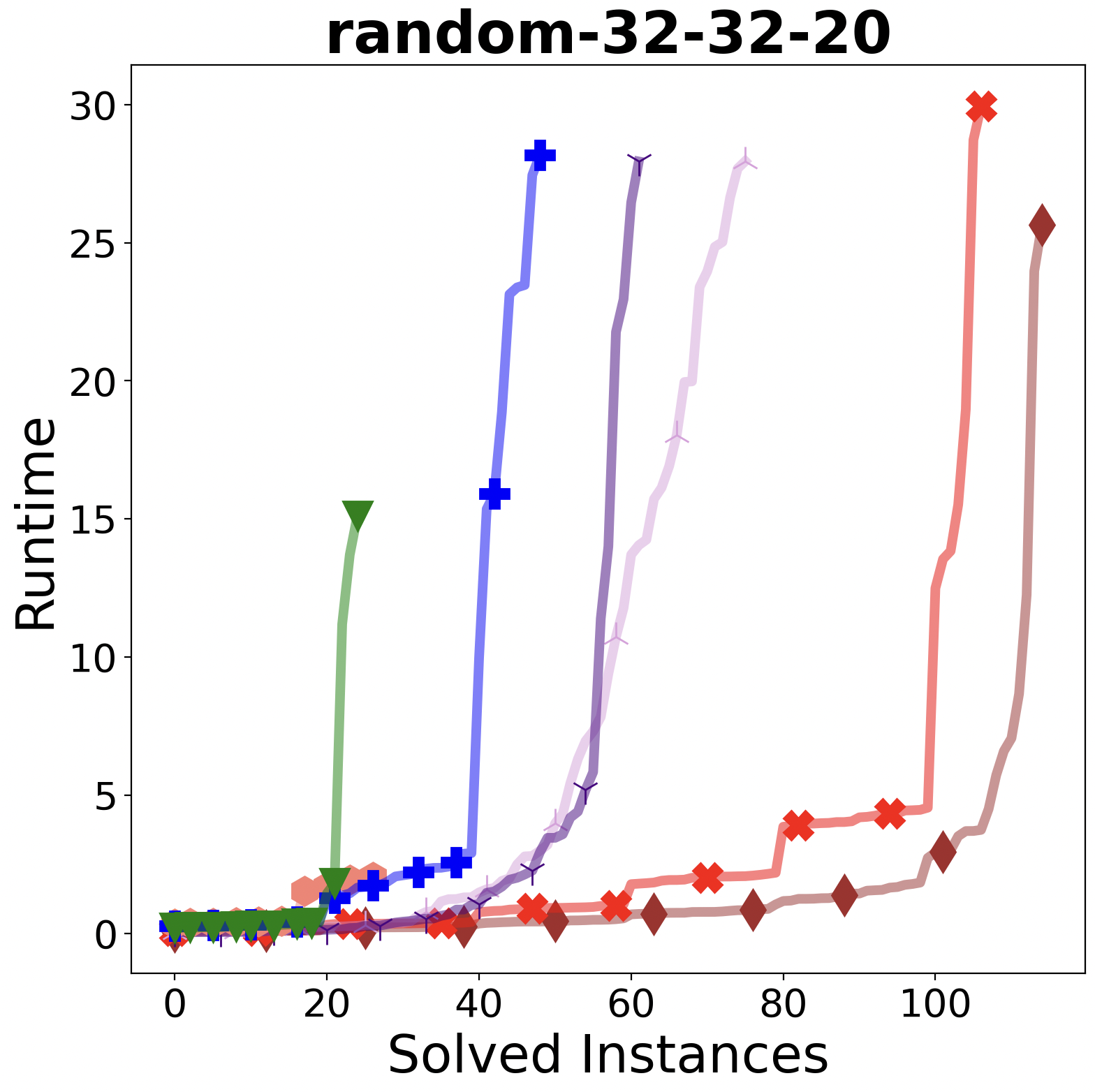}
    \includegraphics[scale=\graphsize]{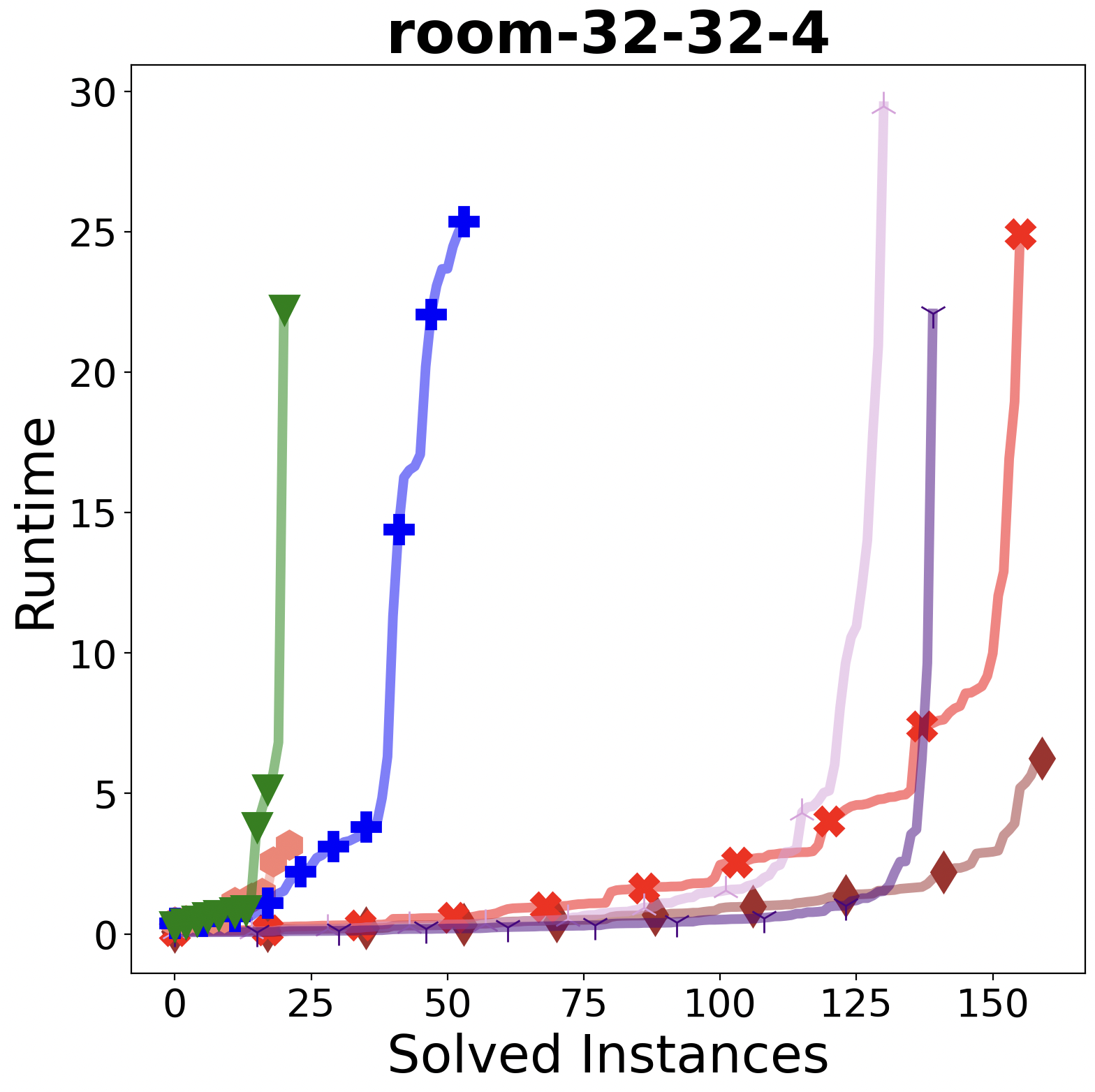}
    \includegraphics[scale=\graphsize]{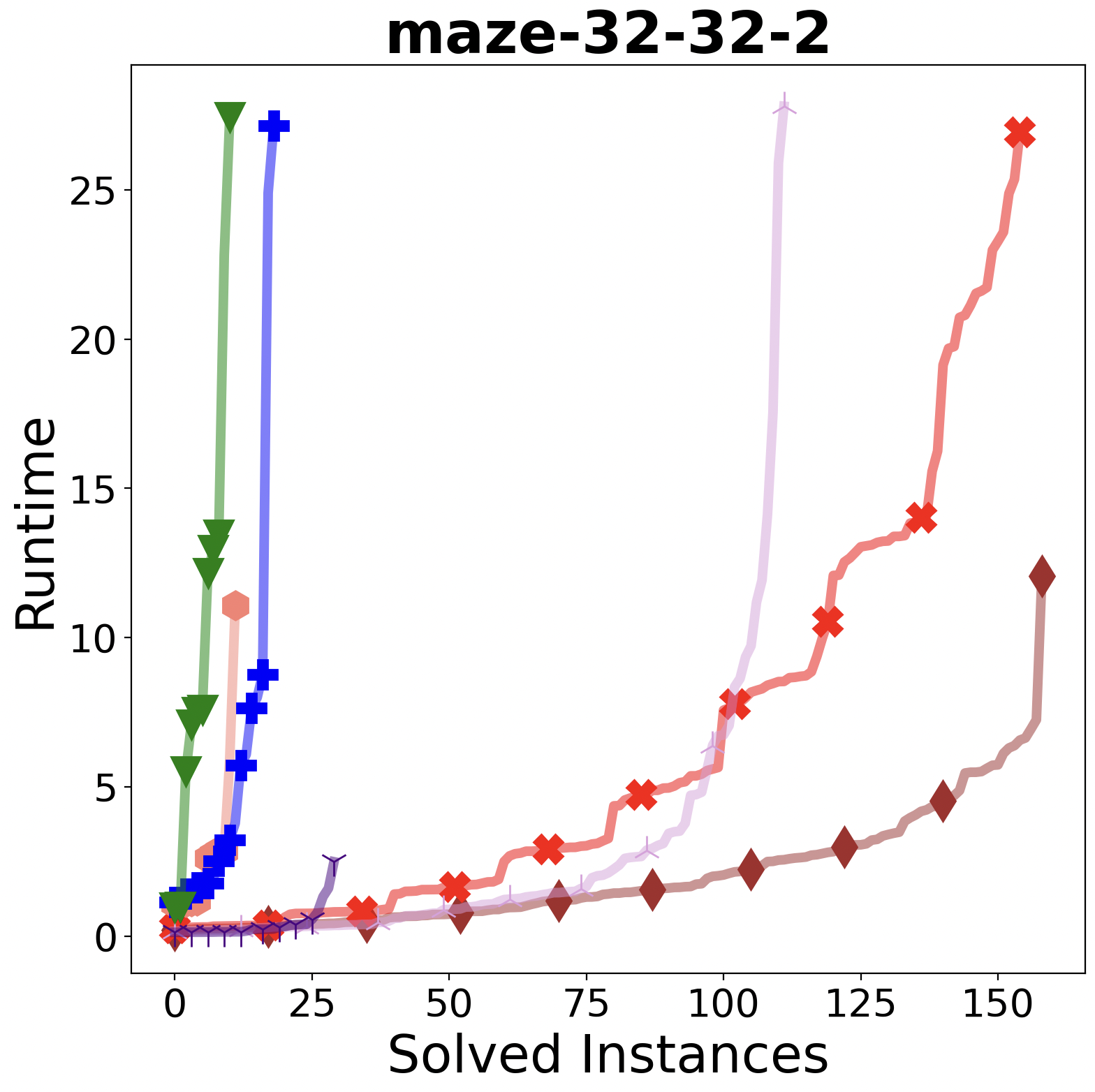}
    \includegraphics[scale=\graphsize]{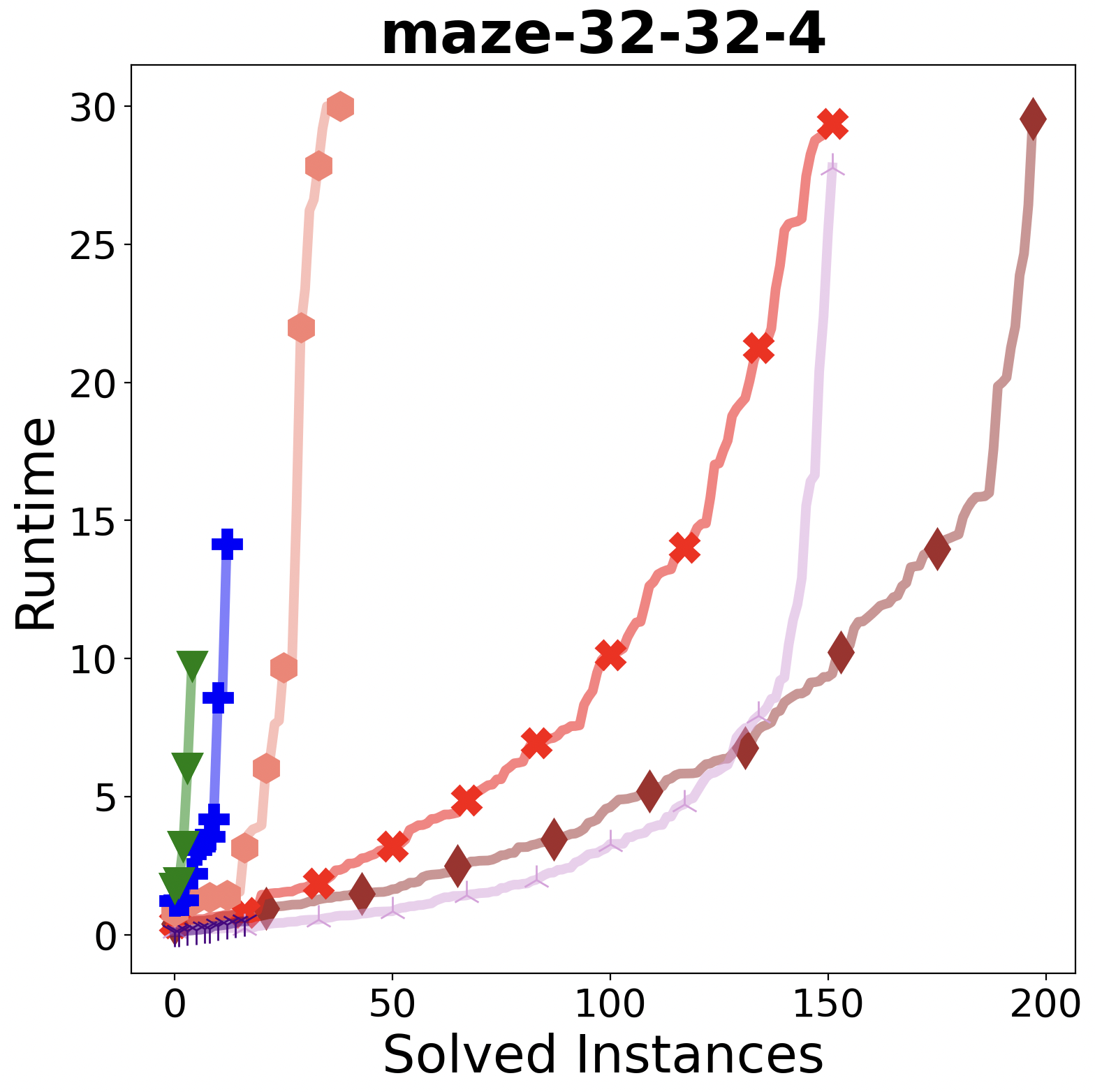}
  \caption{Runtime (Cactus Graph)}
  \label{fig:rt}
\end{figure}
Fig.~\ref{fig:rt} plots the of runtime ($y$-axis) required to solve instances  ($x$-axis) in increasing order of time over all instances that an algorithm solved among all the instances given above, i.e., instances with different number of agents are all included here.
This kind of graph is sometimes referred to as \emph{cactus graph} in literature.
In some cases, LaCAM variants were faster than \cgamapf, such as in \emph{empty-32-32} grid.
Nevertheless, in nearly all other instances, \cgapibt outperformed other algorithms by a large margin.
For example, in \emph{maze-32-32-2}, \cgapibt solved more than 100 instances more than LaCAM$^*$ (who was second best).
This can be explained by the fact that with more dense grids and a rising number of agents, the search tree of LaCAM variants grows rapidly, whereas in our algorithms (\cgamapf and \cgapibt), the execution remains rule-based, and the runtime grows much slower.
In all the grids, \cgamapf was significantly slower than \cgapibt.
% For example, in \emph{room-32-32-2} grid with 450 agents, the \cgapibt algorithm was 4 times faster than \cgamapf.
% Note that the runtime grows polynomially with the number of agents. 

% solution quality: makespan
\begin{figure}[!ht]
    \centering
    \includegraphics[scale=\graphsize]{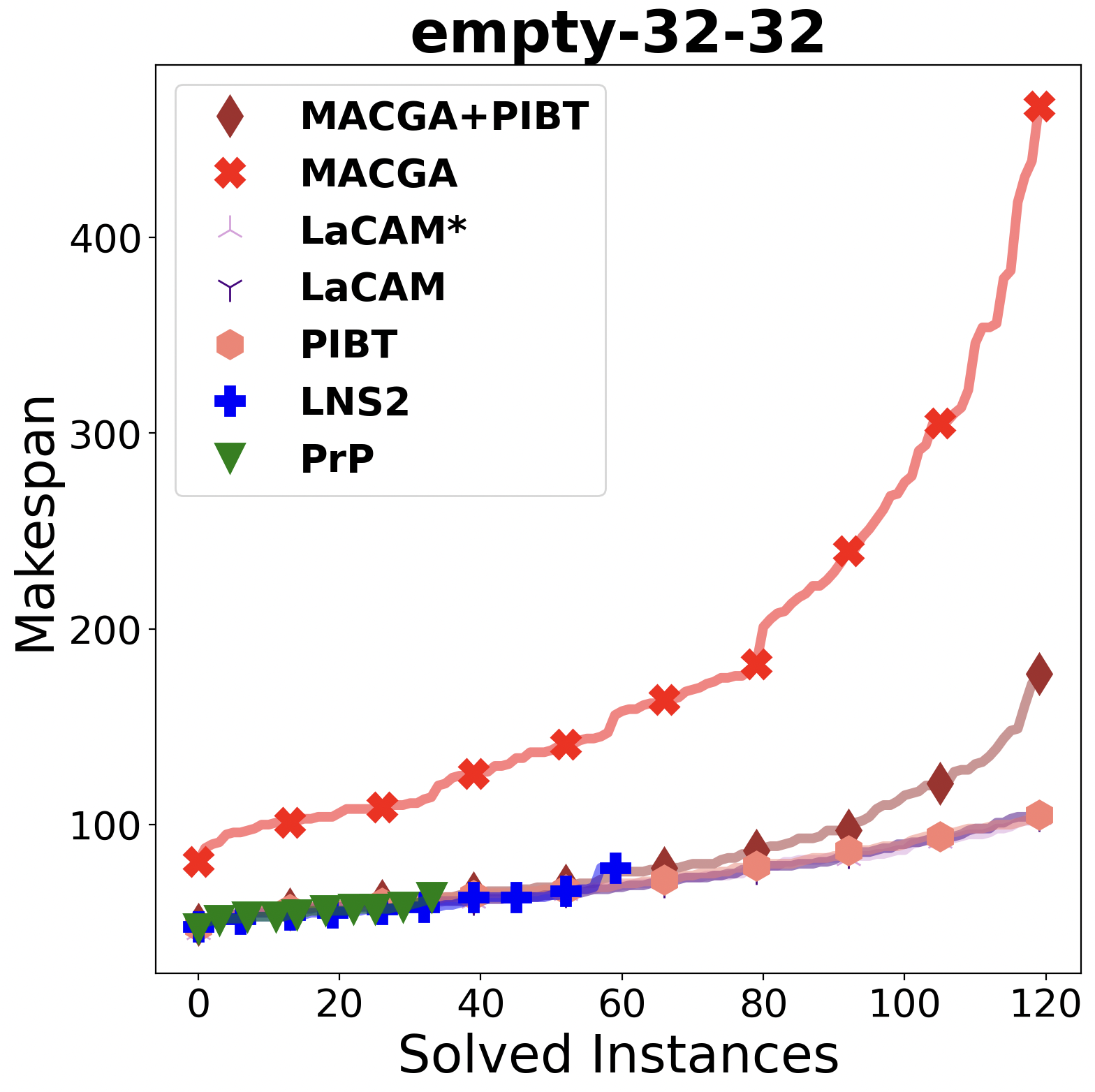}
    \includegraphics[scale=\graphsize]{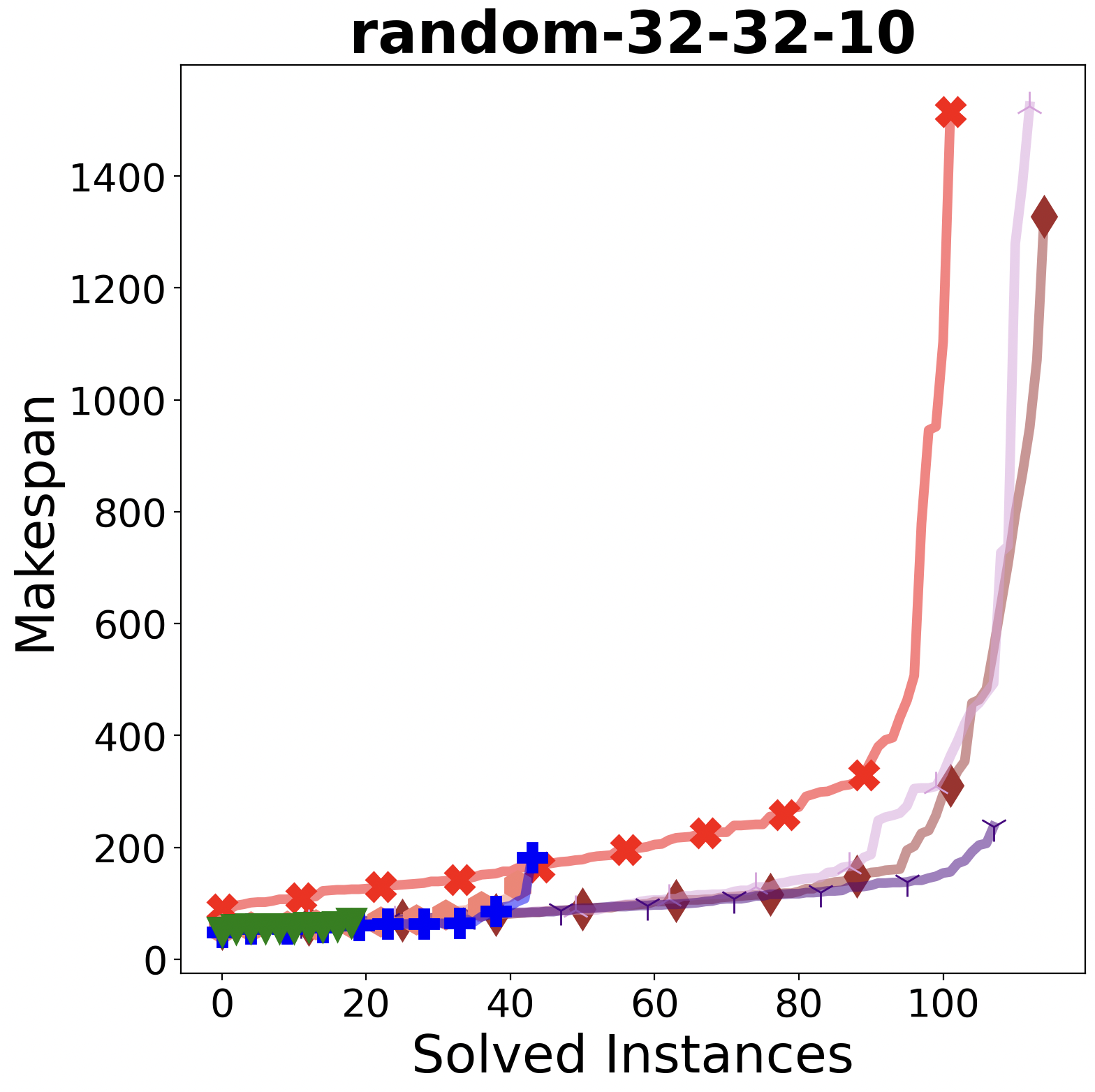}
    \includegraphics[scale=\graphsize]{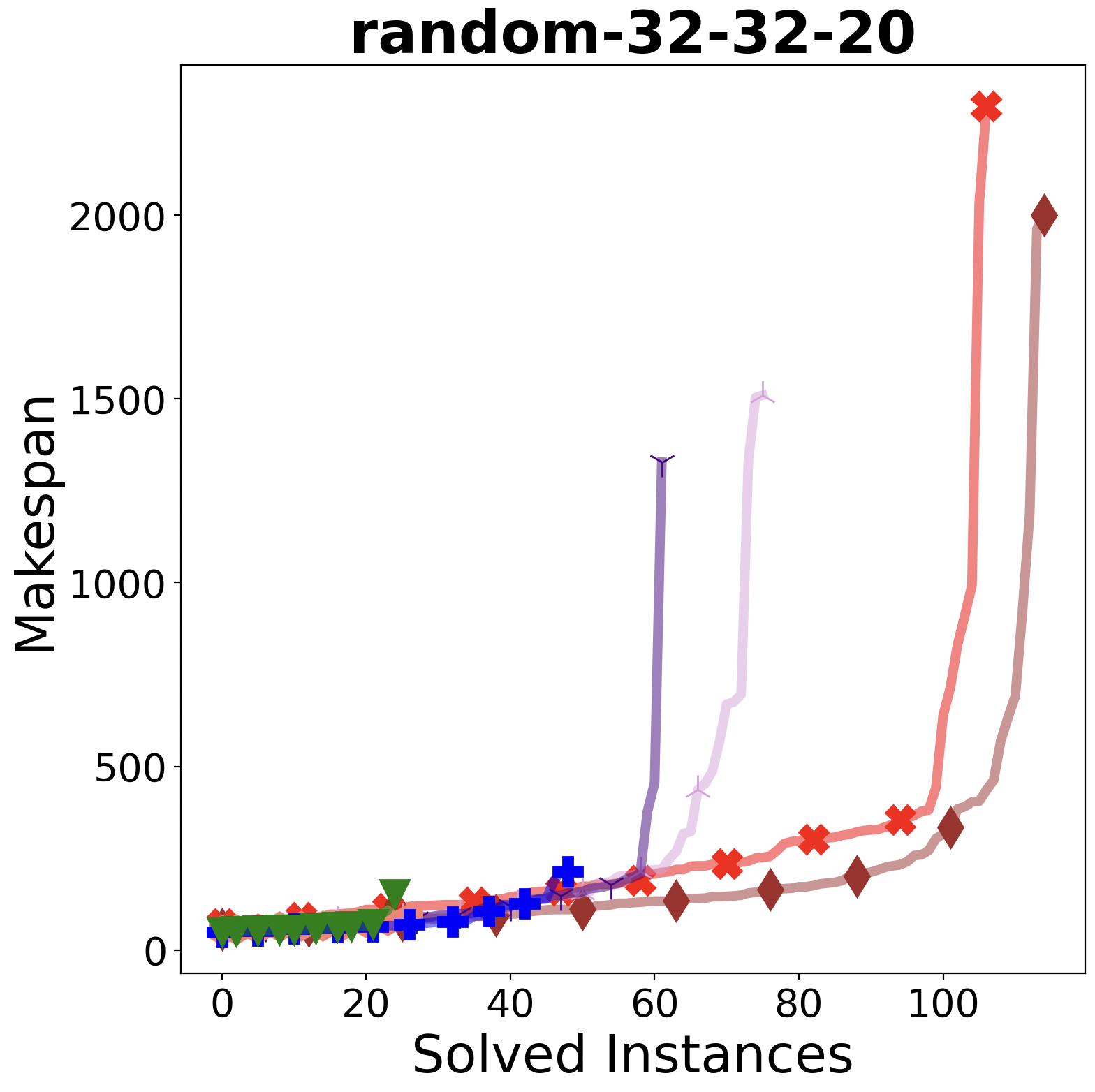}
    \includegraphics[scale=\graphsize]{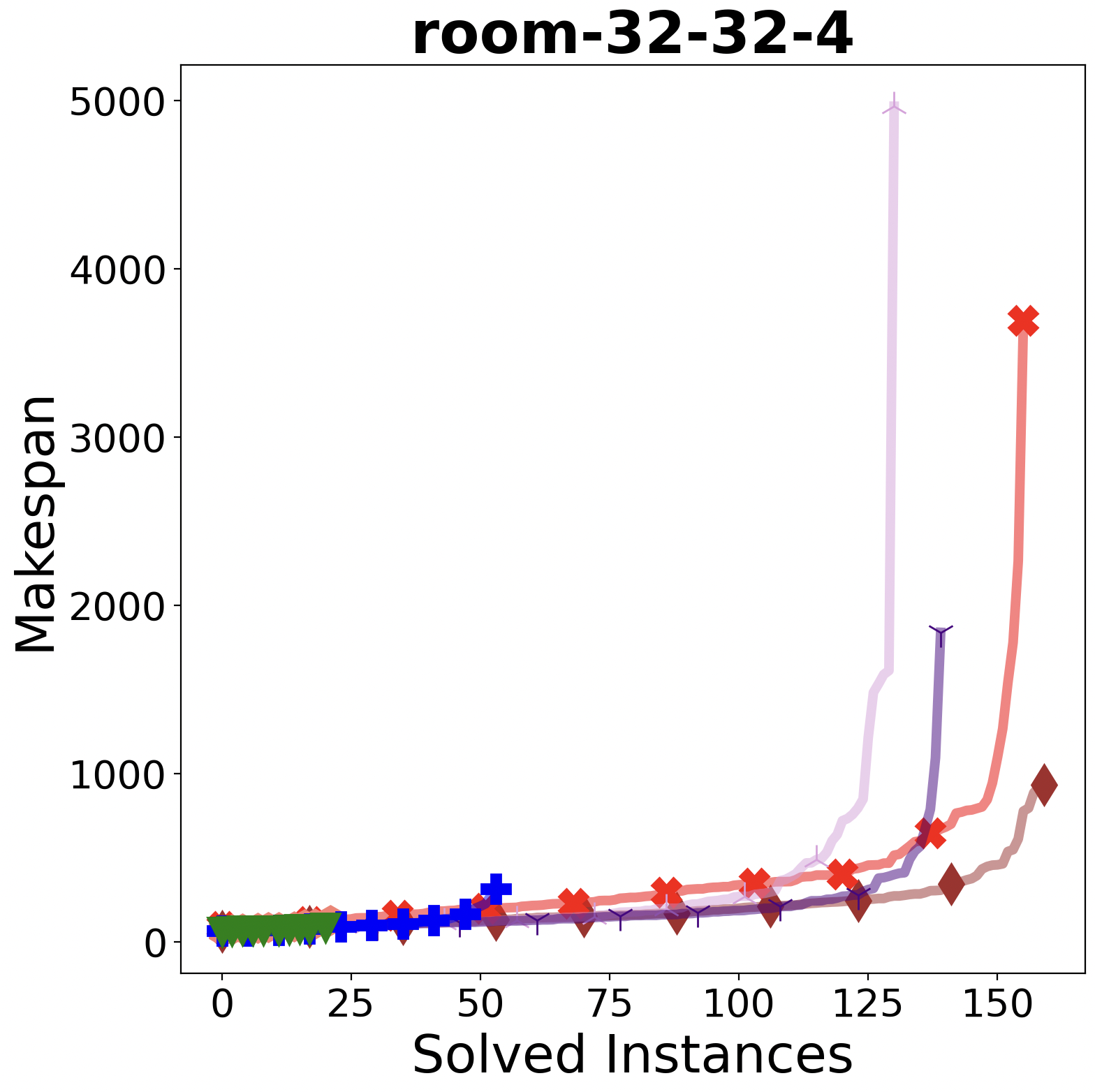}
    \includegraphics[scale=\graphsize]{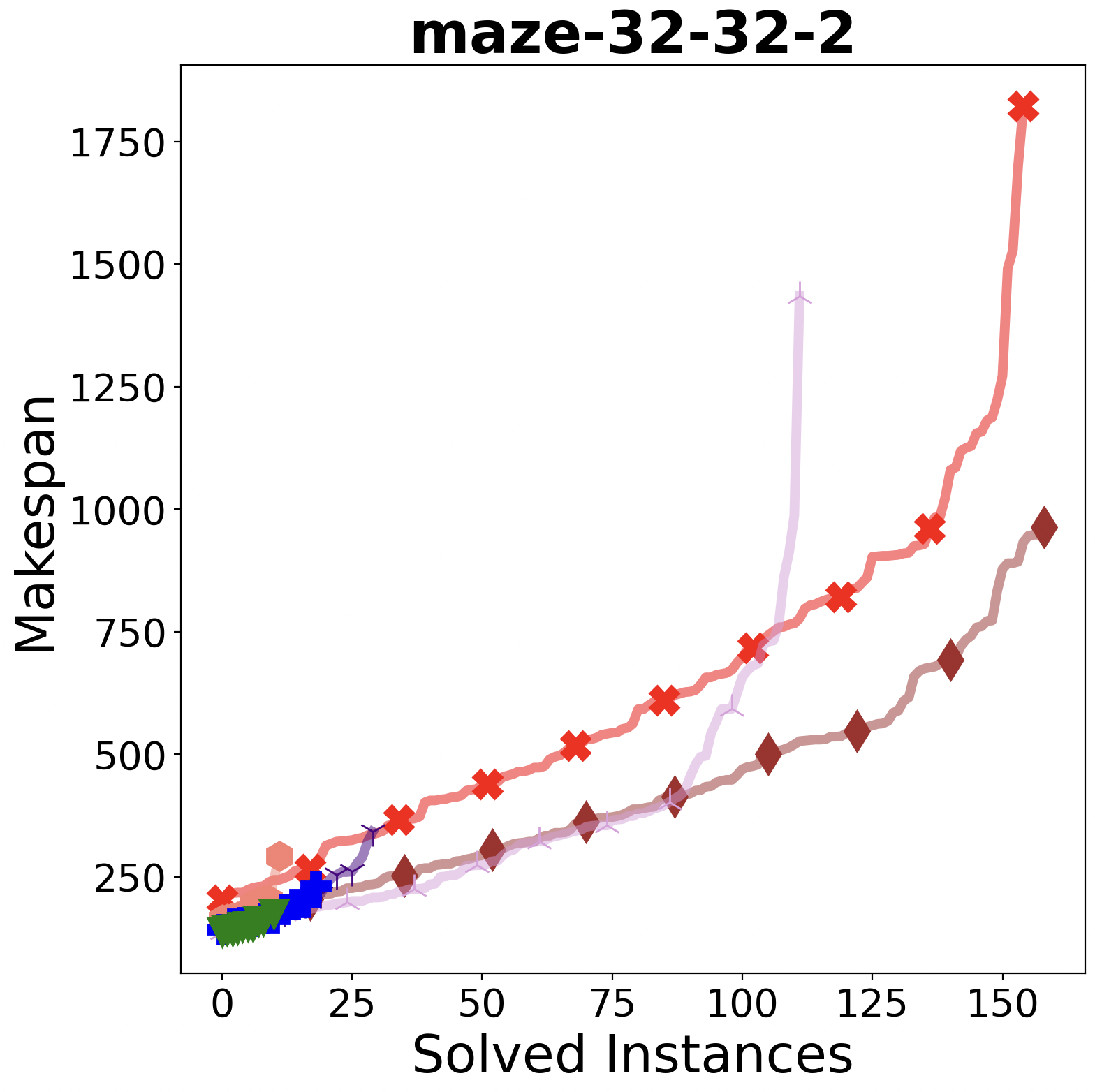}
    \includegraphics[scale=\graphsize]{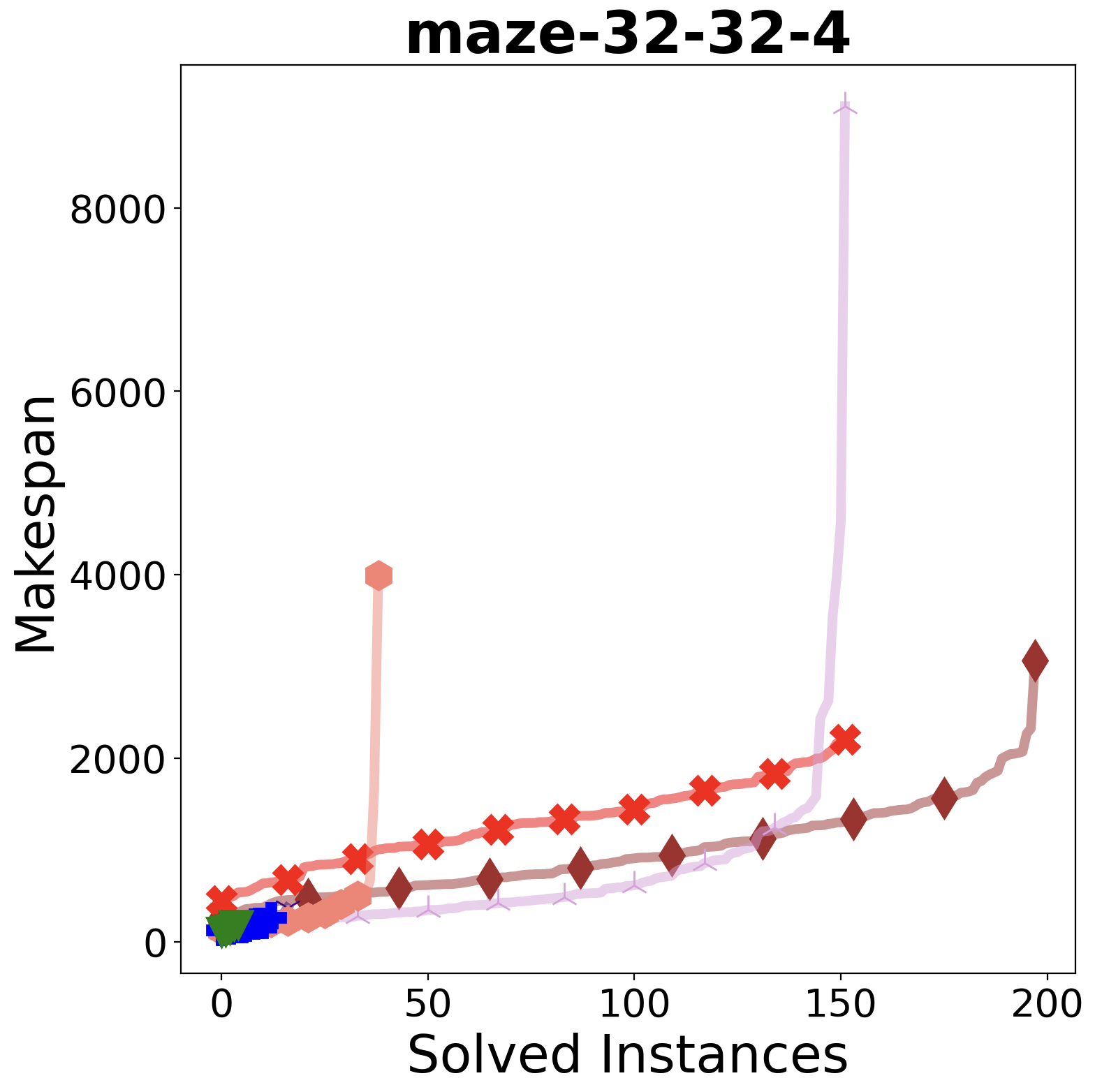}
  \caption{Makespan (Cactus Graph)}
  \label{fig:ms}
\end{figure}
Fig.~\ref{fig:rt} compares the \emph{makespan} obtained by each algorithm by solving instances with a varying number of agents. 
Those graphs are also presented in the form of \emph{cactus graphs}.
The $y$-axis is a makespan value and the $x$-axis designates the solved instances in increasing order of makespan
In most of the cases \cgamapf returns a solution with higher costs, that are substantially higher than the baseline algorithms.
However, \cgapibt preserves much lower costs that are close to the baseline algorithms in the majority of the cases.

% Fig.~\ref{fig:ms} compares the average \emph{makespan} ($y$-axis) obtained by each algorithm (lower is better) per number of agents ($x$-axis). 
% In most of the cases \cgamapf returns a solution with higher costs, that are substantially higher than the baseline algorithms.
% However, \cgapibt preserves much lower costs that are close to the baseline algorithms in the majority of the cases.

% Moreover, in some cases, such as in \emph{random-32-32-20} the \emph{makespan} might be even better for \cgamapf.
% As part of future work, we will focus on improvement regarding solution quality of \cgamapf.

% % solution quality: soc
% \begin{figure*}[!ht]
%     \centering
%     \includegraphics[scale=0.10]{pics/soc_empty.png}
%     \includegraphics[scale=0.10]{pics/soc_rand10.png}
%     \includegraphics[scale=0.10]{pics/soc_rand20.png}
%     \includegraphics[scale=0.10]{pics/soc_room.png}
%     \includegraphics[scale=0.10]{pics/soc_maze2.png}
%     \includegraphics[scale=0.10]{pics/soc_maze4.png}
%   \caption{Sum of Costs}
%   \label{fig}
% \end{figure*}

\section{Conclusion}

We introduced two incomplete rule-based algorithms, \cgamapf and \cgapibt, for solving the Multi-Agent Path Finding (MAPF) problem.
Both algorithms run in polynomial time and guarantee reachability. 
Experimentally, we showed that the proposed approaches solve MAPF problems more efficiently than baseline algorithms in terms of success rate. 
% In the future, we intend to develop algorithms that will ensure completeness and adapt \cgamapf and 
% \cgapibt for lifelong MAPF problems.
In the future, we intend to tailor \cgamapf and 
\cgapibt for lifelong MAPF problems.
Another interesting direction for future work is to adapt the proposed approaches to multi-agent path planning in continuous spaces.
% in terms of solution quality, i.e. \emph{Sum-of-Costs} or \emph{makespan}. 
% Another direction for future work is to adapt \cgamapf for lifelong MAPF problems. 
\section*{Acknowledgements}
This research was partly supported by the Helmsley Charitable Trust through the Agricultural, Biological and Cognitive Robotics Initiative and by the Marcus Endowment Fund, both at Ben-Gurion University of the Negev; by the Binational Science Foundation (BSF) fund \#2022189; and by the Israel Science Foundation (ISF)
grant \#1238/23 to Roni Stern.

\bibliographystyle{named}
\bibliography{ijcai25}

\begin{thebibliography}{}

\bibitem[\protect\citeauthoryear{Abelson \bgroup \em et al.\egroup
  }{1985}]{abelson-et-al:scheme}
Harold Abelson, Gerald~Jay Sussman, and Julie Sussman.
\newblock {\em Structure and Interpretation of Computer Programs}.
\newblock MIT Press, Cambridge, Massachusetts, 1985.

\bibitem[\protect\citeauthoryear{Baumgartner \bgroup \em et al.\egroup
  }{2001}]{bgf:Lixto}
Robert Baumgartner, Georg Gottlob, and Sergio Flesca.
\newblock Visual information extraction with {Lixto}.
\newblock In {\em Proceedings of the 27th International Conference on Very
  Large Databases}, pages 119--128, Rome, Italy, September 2001. Morgan
  Kaufmann.

\bibitem[\protect\citeauthoryear{Brachman and
  Schmolze}{1985}]{brachman-schmolze:kl-one}
Ronald~J. Brachman and James~G. Schmolze.
\newblock An overview of the {KL-ONE} knowledge representation system.
\newblock {\em Cognitive Science}, 9(2):171--216, April--June 1985.

\bibitem[\protect\citeauthoryear{Gottlob \bgroup \em et al.\egroup
  }{2002}]{gls:hypertrees}
Georg Gottlob, Nicola Leone, and Francesco Scarcello.
\newblock Hypertree decompositions and tractable queries.
\newblock {\em Journal of Computer and System Sciences}, 64(3):579--627, May
  2002.

\bibitem[\protect\citeauthoryear{Gottlob}{1992}]{gottlob:nonmon}
Georg Gottlob.
\newblock Complexity results for nonmonotonic logics.
\newblock {\em Journal of Logic and Computation}, 2(3):397--425, June 1992.

\bibitem[\protect\citeauthoryear{Levesque}{1984a}]{levesque:functional-foundations}
Hector~J. Levesque.
\newblock Foundations of a functional approach to knowledge representation.
\newblock {\em Artificial Intelligence}, 23(2):155--212, July 1984.

\bibitem[\protect\citeauthoryear{Levesque}{1984b}]{levesque:belief}
Hector~J. Levesque.
\newblock A logic of implicit and explicit belief.
\newblock In {\em Proceedings of the Fourth National Conference on Artificial
  Intelligence}, pages 198--202, Austin, Texas, August 1984. American
  Association for Artificial Intelligence.

\bibitem[\protect\citeauthoryear{Nebel}{2000}]{nebel:jair-2000}
Bernhard Nebel.
\newblock On the compilability and expressive power of propositional planning
  formalisms.
\newblock {\em Journal of Artificial Intelligence Research}, 12:271--315, 2000.

\end{thebibliography}


\begin{thebibliography}{}

\bibitem[\protect\citeauthoryear{Bart{\'a}k \bgroup \em et al.\egroup }{2019}]{bartak2019multi}
Roman Bart{\'a}k, Ji{\v{r}}{\'\i} {\v{S}}vancara, V{\v{e}}ra {\v{S}}kopkov{\'a}, David Nohejl, and Ivan Krasi{\v{c}}enko.
\newblock Multi-agent path finding on real robots.
\newblock {\em AI Communications}, 2019.

\bibitem[\protect\citeauthoryear{De~Wilde \bgroup \em et al.\egroup }{2013}]{de2013push}
Boris De~Wilde, Adriaan~W Ter~Mors, and Cees Witteveen.
\newblock Push and rotate: cooperative multi-agent path planning.
\newblock In {\em AAMAS}, pages 87--94, 2013.

\bibitem[\protect\citeauthoryear{Felner \bgroup \em et al.\egroup }{2017}]{felner2017search}
Ariel Felner, Roni Stern, Solomon Shimony, Eli Boyarski, Meir Goldenberg, Guni Sharon, Nathan Sturtevant, Glenn Wagner, and Pavel Surynek.
\newblock Search-based optimal solvers for the multi-agent pathfinding problem: Summary and challenges.
\newblock In {\em SoCS}, 2017.

\bibitem[\protect\citeauthoryear{Li \bgroup \em et al.\egroup }{2021}]{li2021anytime}
Jiaoyang Li, Zhe Chen, Daniel Harabor, P~Stuckey, and Sven Koenig.
\newblock Anytime multi-agent path finding via large neighborhood search.
\newblock In {\em IJCAI}, 2021.

\bibitem[\protect\citeauthoryear{Li \bgroup \em et al.\egroup }{2022}]{li2022mapf}
Jiaoyang Li, Zhe Chen, Daniel Harabor, Peter~J Stuckey, and Sven Koenig.
\newblock Mapf-lns2: Fast repairing for multi-agent path finding via large neighborhood search.
\newblock In {\em AAAI}, 2022.

\bibitem[\protect\citeauthoryear{Luna and Bekris}{2011}]{luna2011push}
Ryan~J Luna and Kostas~E Bekris.
\newblock Push and swap: Fast cooperative path-finding with completeness guarantees.
\newblock In {\em IJCAI}, 2011.

\bibitem[\protect\citeauthoryear{Ma \bgroup \em et al.\egroup }{2017}]{ma2017feasibility}
Hang Ma, Jingxing Yang, Liron Cohen, T.~K.~Satish Kumar, and Sven Koenig.
\newblock Feasibility study: Moving non-homogeneous teams in congested video game environments.
\newblock In {\em AIIDE}, 2017.

\bibitem[\protect\citeauthoryear{Morris \bgroup \em et al.\egroup }{2016}]{morris2016planning}
Robert Morris, Corina~S Pasareanu, Kasper~S{\o}e Luckow, Waqar Malik, Hang Ma, TK~Satish Kumar, and Sven Koenig.
\newblock Planning, scheduling and monitoring for airport surface operations.
\newblock In {\em AAAI Workshop: Planning for Hybrid Systems}, 2016.

\bibitem[\protect\citeauthoryear{Okumura \bgroup \em et al.\egroup }{2022}]{okumura2022priority}
Keisuke Okumura, Manao Machida, Xavier D{\'e}fago, and Yasumasa Tamura.
\newblock Priority inheritance with backtracking for iterative multi-agent path finding.
\newblock {\em Artificial Intelligence}, 310:103752, 2022.

\bibitem[\protect\citeauthoryear{Okumura}{2023a}]{okumura2023improving}
Keisuke Okumura.
\newblock Improving lacam for scalable eventually optimal multi-agent pathfinding.
\newblock {\em arXiv}, 2023.

\bibitem[\protect\citeauthoryear{Okumura}{2023b}]{okumura2023lacam}
Keisuke Okumura.
\newblock Lacam: Search-based algorithm for quick multi-agent pathfinding.
\newblock In {\em AAAI}, volume~37, pages 11655--11662, 2023.

\bibitem[\protect\citeauthoryear{Pertzovsky \bgroup \em et al.\egroup }{2024}]{arseni2024corr}
Arseni Pertzovsky, Roni Stern, and Roie Zivan.
\newblock Cga: Corridor generating algorithm for multi-agent environments.
\newblock In {\em IROS}. IEEE, 2024.

\bibitem[\protect\citeauthoryear{Salzman and Stern}{2020}]{salzman2020research}
Oren Salzman and Ron~Zvi Stern.
\newblock Research challenges and opportunities in multi-agent path finding and multi-agent pickup and delivery problems blue sky ideas track.
\newblock In {\em AAMAS}, 2020.

\bibitem[\protect\citeauthoryear{Sharon \bgroup \em et al.\egroup }{2015}]{CBS_2015}
Guni Sharon, Roni Stern, Ariel Felner, and Nathan~R Sturtevant.
\newblock Conflict-based search for optimal multi-agent pathfinding.
\newblock {\em Artificial Intelligence}, 2015.

\bibitem[\protect\citeauthoryear{Silver}{2005}]{prp_2015}
David Silver.
\newblock Cooperative pathfinding.
\newblock In {\em AIIDE}, 2005.

\bibitem[\protect\citeauthoryear{Stern \bgroup \em et al.\egroup }{2019}]{stern2019mapf}
Roni Stern, Nathan~R. Sturtevant, Ariel Felner, Sven Koenig, Hang Ma, Thayne~T. Walker, Jiaoyang Li, Dor Atzmon, Liron Cohen, T.~K.~Satish Kumar, Eli Boyarski, and Roman Bartak.
\newblock Multi-agent pathfinding: Definitions, variants, and benchmarks.
\newblock In {\em SoCS}, pages 151--158, 2019.

\bibitem[\protect\citeauthoryear{Wurman \bgroup \em et al.\egroup }{2008}]{wurman2008coordinating}
Peter~R Wurman, Raffaello D'Andrea, and Mick Mountz.
\newblock Coordinating hundreds of cooperative, autonomous vehicles in warehouses.
\newblock {\em AI magazine}, 29(1):9--9, 2008.

\end{thebibliography}

\end{document}